\documentclass[preprint,12pt]{elsarticle}

\usepackage{cite}
\usepackage{booktabs}
\usepackage{amsmath,amssymb,amsfonts}
\usepackage{algorithmic}
\usepackage{graphicx}
\usepackage{epstopdf}
\usepackage{textcomp}
\usepackage{xcolor}
\usepackage{colortbl}
\usepackage{arydshln}
\usepackage{multirow}
\usepackage{multicol}
\usepackage{subfigure}
\usepackage{hyperref}
\usepackage{url}
\usepackage{xspace}
\usepackage{ulem}
\usepackage{array}
\usepackage{ragged2e}                  
\usepackage[scale=0.85]{geometry}
\usepackage{threeparttable}
\usepackage{breakurl}

\def\BibTeX{{\rm B\kern-.05em{\sc i\kern-.025em b}\kern-.08em
    T\kern-.1667em\lower.7ex\hbox{E}\kern-.125emX}}

\newcommand{\convec}[1]{\mathbf{v}(#1)}
\newcommand{\elevec}[1]{\mathbf{l}(#1)}

\newcommand{\newsemb}[1]{\mathbf{n}}

\newcommand{\ourmodel}{D-HAN}
\journal{Nuclear Physics B}

\begin{document}

\begin{frontmatter}

\title{Dynamic Hierarchical Attention Network for News Recommendation}

\author[inst1]{Qinghua Zhao\corref{mycorrespondingauthor}}
\cortext[mycorrespondingauthor]{Corresponding author}
\ead{zhaoqh@buaa.edu.cn}

% \author[inst2]{Xu Chen\corref{mycorrespondingauthor}}
% \cortext[mycorrespondingauthor]{Corresponding author}
% \ead{xu.chen@ruc.edu.cn}

% \author[inst1]{Hui Zhang}
% \ead{zhangh17@buaa.edu.cn}

\affiliation[inst1]{organization={SKLSDE Lab, Beihang University},%Department and Organization
            addressline={37 Xueyuan Road}, 
            city={Beijing} }

% \affiliation[inst2]{organization={  Beijing Key Laboratory of Big Data Management and Analysis Methods, Gaoling School of Artificial Intelligence},
%             addressline={Renmin University of China}, 
%             city={Beijing} 
%             }  
% \affiliation[ucph]{organization={Department of Computer Science, University of Copenhagen},
%     city={Lyngbyvej 2, Copenhagen Ø},
%     postcode={2100}, 
%     country={Denmark}}
 % \ead{\{zhaoqh, zhangh17, mashuai\}@buaa.edu.cn. xu.chen@ruc.edu.cn. }

\begin{abstract}
News recommendation models often fall short in capturing users' preferences due to their static approach to user-news interactions. To address this limitation, we present a novel dynamic news recommender model that seamlessly integrates continuous time information to a hierarchical attention network that effectively represents news information at the sentence, element, and sequence levels. Moreover, we introduce a dynamic negative sampling method to optimize users' implicit feedback. To validate our model's effectiveness, we conduct extensive experiments on three real-world datasets. The results demonstrate the effectiveness of our proposed approach. 
% Our source code is available at \url{https://github.com/lshowway/D-HAN}.
\end{abstract}


\end{frontmatter}

\section{Introduction}
News recommendation refers to the process of automatically suggesting or recommending news articles to users that they are likely to find interesting or valuable. In this process, it takes into account various  factor such as the user's reading history, explicit feedback (e.g., ratings or likes), implicit feedback (e.g., clicks or dwell time), demographic information, and contextual data, just to name a few. To implement this, natural language processing (NLP) and information retrieval (IR) are the commonly used techniques, and many methods have been proposed \citet{dwivedi2016survey,zheng2018drn,LATIFI2021291,wu2020mind,KHAN202269,wu2023personalized,PHAM2023105}.

In the field of news recommendation, an important observation is that people's news-reading behaviors are not independent~\citet{DBLP:conf/recsys/GarcinDF13,zhang2022metonr}. Previous interactions with news articles have a significant influence on subsequent reading choices. As a result, numerous news recommendation models have been developed to capture people's sequential reading patterns~\citet{DBLP:conf/cikm/ParkLC17, DBLP:conf/cikm/KhattarKV018, DBLP:conf/cikm/KhattarKV018a}. Furthermore, news articles are composed of multiple sentences, each playing a distinct role within the article.
Additionally, news articles consist of essential components known as news elements, commonly referred to as the five W's and one H (5W1H): ``who, when, where, what, why, how''~\citet{li2007flexible}. These elements explicitly describe critical information in the news and are followed by the world press as a basic principle of news writing. 
Therefore, user-news interactions involve not only the news article itself (including its sentences), but also news elements and user interactions. Given the diverse information available in news reading scenarios, it is crucial to consider them at the sentence-, element-, and sequence-levels. 
Moreover, the current period and the time difference since the last clicked news significantly impact subsequent news recommendations. For example, users may browse stock market news at around 10 am on weekdays and constellations news at midnight, emphasizing the need to incorporate time attributes in news recommendations.

Motivated by the aforementioned observations, we introduce \textbf{\ourmodel}, a \emph{D}ynamic \emph{H}ierarchical \emph{A}ttention \emph{N}etwork for news recommendation. Our model comprises a hierarchical network with two layers: the first layer focuses on processing news article information, including sentences and elements, while the second layer handles news sequential information, i.e., the user's clicked history of news articles. Furthermore, we consider dynamic information such as absolute time (news clicked timestamps) and relative time (the time difference between two clicked news). To guide the optimization process, we employ a novel dynamic negative sampling approach.

Building upon the work of \citet{zhang2019dynamic}, our contributions can be summarized as follows: a) retaining the main hierarchical block while replacing the convolutional neural network with the Transformer \citet{vaswani2017attention}, b) incorporating dynamic information, including absolute and relative time, and c) designing a novel dynamic negative sampling method to guide the optimization process.

\section{Related Work}
News recommendation has been a topic of significant interest, aiming to provide personalized news articles for users. Traditional methods for news recommendation can be categorized into content-based, collaborative filtering, and hybrid approaches. Content-based methods recommend news articles based solely on content similarity~\citet{DBLP:conf/www/LvMKZWC11, DBLP:conf/www/HsiehYWNE16}. Collaborative filtering methods leverage user feedback on news articles to make recommendations~\citet{DBLP:conf/www/DasDGR07}, but they often suffer from cold-start problems. Hybrid methods combine both strategies to achieve better recommendation performance~\citet{DBLP:conf/sigir/LiWLKP11, DBLP:journals/ipm/LvMZ17}. In recent studies, attention networks have been applied to model user representations in news recommendation~\citet{DBLP:conf/www/WangZXG18, DBLP:conf/kdd/ZhouZSFZMYJLG18, lian2018towards}.

With the remarkable success of the Transformer model \citet{vaswani2017attention}, it has been widely employed in news recommendation research \citet{10.1145/3543507.3583383, 10.1145/3459637.3482401, 10.1145/3511808.3557335, HUANG2023118943, liu2023perconet}. BERT4Rec \citet{sun2019bert4rec} concatenates all history news articles into a document as input for the Transformer encoder. NRMS \citet{wu2019neural} proposes a news encoder and a user encoder, where multi-head self-attention is utilized to learn representations. \citet{HUANG2022119} models the inherent relatedness between previously clicked news and candidate news. \citet{ZHU202233} explores the integration of dynamic items, such as social information of users, to improve user representations. \citet{10.1145/3511708} presents a graph neural news recommendation model that incorporates existing and potential user interests. Additionally, \citet{10.1145/3555373} proposes an approach that integrates both user-dependent preference and user-independent timeliness in news recommendation. \citet{liu2023perconet} focuses on representing users better by incorporating entities recognized in the history news.

Dynamic recommendation takes into account the time-dependent effect in the recommendation process, providing deeper insights into users' behaviors. For instance, \citet{DBLP:conf/sigir/QuanYuan13} observes that individuals tend to visit different locations at different times of the day and utilizes absolute time for location recommendation. To capture the dynamics in sequence recommendation, Time-LSTM \citet{DBLP:conf/ijcai/ZhuLLWGLC17} recognizes the significance of time intervals between users' actions and integrates time gates into LSTM to incorporate time interval information within the RNN framework. Another approach, TiSASRec \citet{li2020time}, models user interaction sequences by considering different time intervals computed from the timestamps of consecutive clicked articles. In TiSASRec, the relative time interval is modeled as a relation matrix capturing the relationships between any two items.

\section{{\ourmodel} Model}
\subsection{Preliminaries}
\begin{figure*}[t]
\centering
\includegraphics[width=0.6\textwidth]{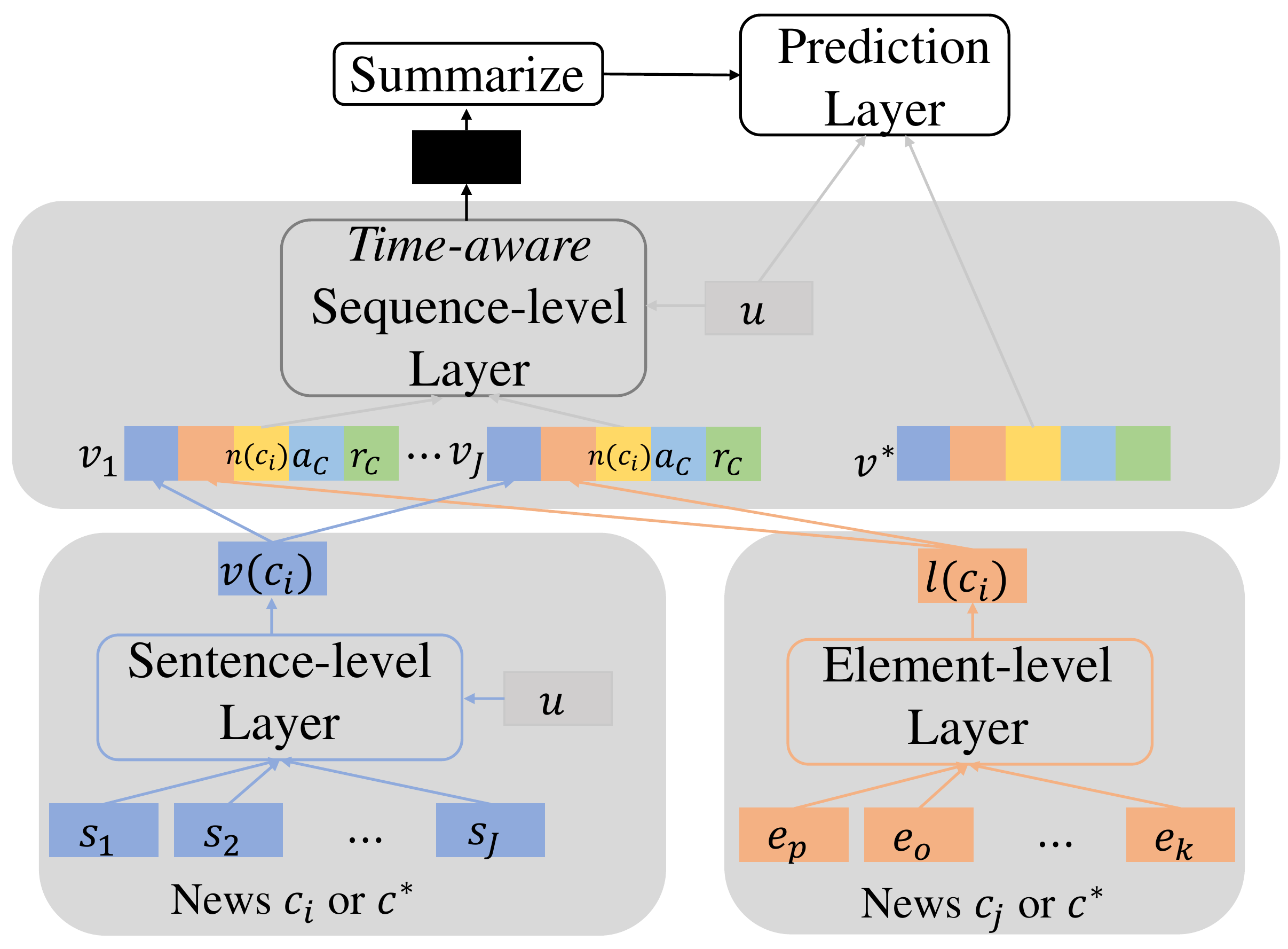}
\caption{The architecture of \ourmodel~model. It has three main components including sentence-level, element-level, and time-aware sequence-level attention layer. To train the model, dynamic negative sampling method is also applied. }
\label{framework}
\end{figure*}

News recommendation can be defined as estimating the click rate of the next news article for a user based on their previously read news articles. Formally, we consider a user's sequence of the most recent $L$ news articles as $\mathcal{C}=[c_1,c_2,...,c_L]$, where $c_i$ represents a piece of news and $L$ is the number of news articles considered for click rate estimation. Each news article $c_i$ consists of a sequence of sentences, denoted as $[s_{i1},s_{i2},...,s_{iJ}]$, where $s_{ij}$ represents the $j$-th sentence in news article $c_i$, and $J$ is the maximum number of sentences considered. 
Additionally, we consider news elements including \textit{person}, \textit{organization}, \textit{time}, \textit{location}, and \textit{keywords}. Hence, each news article $c_i$ is represented by a set of elements, namely ${e_{ip},e_{io},e_{it},e_{il},e_{ik}}$, corresponding to the defined elements. Given the news sequence $\mathcal{C}$ and a candidate news article $c^*$, our objective is to predict the click rate of $c^*$. Figure~\ref{framework} shows the framework of our model.
%%%%%%%%%%%%%%%%%%%%%%%%%%%%%%%%%%
%%%%%%%%%%%%%%%%%%%%%%%%%%%%%%%%%%

\subsection{Sentence-level Layer}\label{sen_attn_model}
The sentence-level attention layer aims to discriminate various influences of sentences. When predicting candidate news $c^*$, sentences of news $c_i$ unequally affect user's choices. To implement this, we first represent sentences in a given piece of news $c_i$. The vector $\convec{s_{ij}}$ of sentence $s_{ij}$ is embedded in a $d$-dimensional space $\mathbb{R}^d$, and the vector $\convec{c_i} \in \mathbb{R}^{K \times d}$ of news $c_i$ is stacked by $\convec{s_{ij}}$, and the vector $\convec{c^*}\in\mathbb{R}^d$ of candidate news $c^*$ is calculated by averaging the sentences vectors.
Moreover, user identity information is learned by an embedding in a $d$-dimensional space, i.e., $\mathbf{u} \in\mathbb{R}^d$ for user $u$, which is randomly initialized and updated during training.
To model the interactions among news $c_{i}$ and candidate news $c^*$, we concatenate $\mathbf{u}$, $\convec{c_{i}}$ and $\convec{c^*}$ along the first dimension as $[\mathbf{u} \convec{{c}_{i}} \convec{{c}^*}] \in\mathbb{R}^{(K+2) \times d}$, and adopt scaled dot-product attention \citet{vaswani2017attention} to learn its representation. The self attention weights among $c_{i}$ and $c^*$ are:

\begin{equation}\label{eq:sen_vec_1}
{\beta}_{i} = \frac{[\mathbf{u} \convec{{c}_{i}}\convec{{c}^*}]\mathbf{W}_1    \cdot (\mathbf{W}_2[\mathbf{u} \convec{{c}_{i}}\convec{{c}^*}]^T)}{\sqrt{d}},
\end{equation}

where $\mathbf{W}_{1}\in\mathbb{R}^{d\times d}$, $\mathbf{W}_{2} \in\mathbb{R}^{d \times d}$ are the parameters of model, $\cdot$ denotes the dot product operation,  $\sqrt{d}$ is a scale factor to prevent dot products growing large in magnitude, $\beta_i \in \mathbb{R}^{(K+2) \times (K+2)}$. Note that  
we omit the bias for brevity here.
These self attention weights are further normalized by softmax function, which can be interpreted as the content relevance among sentence $s_{ij}$ and candidate news $c^*$, user embedding $\mathbf{u}$.
With these weights, we compute the vector $\convec{c_i}$ of news $c_i$ with respect to candidate news ${c}^{*}$ as the dot product between ${\beta}_{i}$ and linearly transformed  $[\mathbf{u} \convec{{c}_{i}}\,\convec{{c}^*}]$ by model parameter $\mathbf{W_{3} \in \mathbb{R}^{d \times d}}$, that is, 
\begin{equation}
    \convec{{c}_{i}}= {\beta}_{i} \cdot  [\mathbf{u} \convec{{c}_{i}} \convec{{c}^*}] \mathbf{W_{3}} \in\mathbb{R}^{(K+2) \times d}
\end{equation}

\subsection{Element-level Layer}\label{ele_attn_model}
Given a pair of news $c_i$ and candidate news $c^*$, different elements play a different role in users' decisions, and the goal of the element-level attention layer is to discriminate various impacts of other elements. With the named entity recognition and keywords extraction modules of NLP tools, we can extract elements we define, i.e., person, organization, time, location and keywords.
Each element is extracted in the form of one or more words.
In this paper, each word is embedded in a $d$-dimensional space $\mathbb{R}^d$.
For instance, let $\mathbf{l}{c_i} = \{\mathbf{l}{e_{ip}}, \mathbf{l}{e_{io}}, \mathbf{l}{e_{it}}, \mathbf{l}{e_{il}}, \mathbf{l}{e_{ik}}\} \in \mathbb{R}^{5 \times d}$, where $\mathbf{l}{e_{ip}} \in \mathbb{R}^d$ refers to element \textit{person} $p$'s vector in news $c_i$, which is obtained by averaging the vectors of words that represent $e_{ip}$.
We concatenate the corresponding element vector of news $c_i$ and $c^*$ along $d$-dimension as $[\mathbf{l}{c_i} \mathbf{l}{c^*}] \in \mathbb{R}^{5 \times 2d}$, and then to smooth their internal differences, we apply three one-layer feed-forward neural networks to transform them into $d$-dimension space:

\begin{equation} \label{eq:ele_vec_1}
\begin{aligned}
    q_{c_{i}}= [\mathbf{l}{c_i} \mathbf{l}{c^*}] \mathbf{W_{4}}, \\
    k_{c_{i}}= [\mathbf{l}{c_i} \mathbf{l}{c^*}] \mathbf{W_{5}}, \\
    v_{c_{i}}= [\mathbf{l}{c_i} \mathbf{l}{c^*}] \mathbf{W_{6}},
\end{aligned}
\end{equation}

where $\mathbf{W_{4}} \in \mathbb{R}^{2d \times d}$, $\mathbf{W_{5}} \in \mathbb{R}^{2d \times d}$ and $\mathbf{W_{6}} \in \mathbb{R}^{2d \times d}$ are model parameters, $q_{c_{i}}$, $ k_{c_{i}}$ and $v_{c_{i}}$ represent different linear transformations of $[\mathbf{l}{c_i} \mathbf{l}{c^*}]$, which will be used to compute self-attended vector representation.
To model the internal and external interactions among element vector of $c_i$ and $c^*$, we take dot product between $q_{c_i}$ and $k_{c_{i}}$ to get the raw self-attention scores:
\begin{equation} \label{eq:ele_vec_2}
    {\gamma}_{i}= \frac{q_{c_i} \cdot {k_{c_{i}}}^T}{\sqrt{d}} \in \mathbb{R}^{5 \times 5}, 
\end{equation}
where $\sqrt{d}$ is a scale factor, and the raw self-attention scores are normalized by softmax function to probabilities followed by a dropout operation on entire elements to attend to.
Here self-attention scores ${\gamma}_{i}$ can be interpreted as the relevance of elements under the influence of both historical news $c_i$ and candidate news $c^*$. With these weights, we compute the element vector $\elevec{c_i}$ of news $c_i$ as dot product between ${\gamma}_{i}$ and $v_{c_{i}}$: 
\begin{equation}
    \elevec{c_i} = {\gamma}_i \cdot v_{c_{i}} \in \mathbb{R}^{5 \times d}
\end{equation}

\subsection{Time-aware Sequence-level Layer}
Given a news-reading sequence $\mathcal{C}$, news articles unequally influence whether he reads candidate news $c^*$ or not, and different periods are also crucial for making decisions. 
To preserve the news identity information, we learn a news id embedding in a $d$-dimensional space for each news according to its id, i.e., $\mathbf{n} (c_i) \in\mathbb{R}^d$ for news $c_i$. News id embeddings are randomly initialized and automatically learned in the training phase. 
For news $c_i$, we concatenate its sentence, element and identity embedding along this $d$-dimension to obtain representation $\mathbf{x'}_{c_i} = [\,\convec{c_i}\,\elevec{c_i}\, \mathbf{n}(c_i)] \in \mathbb{R}^{3d}$.

An observation is that users tend to read different news at different periods with a distinct time difference from the last clicked time. To model this dynamics of news recommendation, we propose to model news content information and timestamp information simultaneously. Specifically, we adopt the clicked timestamp, including year, month, week, day, hour, minutes, we denote it as absolute time. Moreover, we also use the absolute time difference between news $c_i$ and the last clicked news, we call it relative time. To preserve the time information, we learn $d$-dimensional absolute time embedding and relative time embedding for news sequence $\mathcal{C}$, i.e., $\mathbf{a}_{\mathcal{C}} \in \mathbb{R}^{(L+1) \times d}$ for absolute time and $\mathbf{r_{\mathcal{C}}} \in \mathbb{R}^{L \times d}$ for relative time interval.

\noindent
$\bullet$ \textbf{Embed with $\mathbf{r_{\mathcal{C}}}$.}
Given a news sequence $\mathcal{C}$, its representation $\mathbf{x'}_{\mathcal{C}}$ is concatenated with its corresponding relative time interval embedding $\mathbf{r_{\mathcal{C}}}$ and candidate news representation $\mathbf{x^*}$ as  $\mathbf{z}_{\mathcal{C}} = [\mathbf{x'}_{\mathcal{C}} \mathbf{r_{\mathcal{C}}} \mathbf{x}^*] \in \mathbb{R}^{L \times 7d}$, where $\mathbf{x'}_{\mathcal{C}} = \{ \mathbf{x'}_{c_1}, \mathbf{x'}_{c_2}, ..., \mathbf{x'}_{c_L} \} \in \mathbb{R}^{L \times 3d}$. The time-aware representation of candidate news is $\mathbf{z^*} = \mathbf{x^*}$.

\noindent
$\bullet$ \textbf{Embed with $\mathbf{a_{\mathcal{C}}}$.}
History news representation and candidate news representation are concatenated with their corresponding absolute time embedding as $\mathbf{z}_{\mathcal{C}} = [\mathbf{x'}_{\mathcal{C}} \mathbf{a}_{\mathcal{C}}^{1:L}] \in \mathbb{R}^{L \times 4d}$, $\mathbf{z^*} = [\mathbf{x^*} \mathbf{a}_{\mathcal{C}}^{L+1}] \in \mathbb{R}^{4d}$, respectively. 

\noindent
$\bullet$ \textbf{Embed with both $\mathbf{r_{\mathcal{C}}}$ and $\mathbf{a_{\mathcal{C}}}$.} 
First, to use the absolute timestamp and relative time interval, we concatenate $\mathbf{x'_{\mathcal{C}}}$, $\mathbf{a_{\mathcal{C}}}^{1:L}$ and $\mathbf{r_{\mathcal{C}}}$ as $\mathbf{\mathbf{z}_{\mathcal{C}}} = [\mathbf{x'}_{\mathcal{C}} \mathbf{a}_{\mathcal{C}}^{1:L} \mathbf{r_{\mathcal{C}}}] \in \mathbb{R}^{L \times 5d}$, 
Then, to embed time embedding into candidate news representation, we concatenate the representation $\mathbf{x^*}$ of candidate news $c^*$ and the representation $\mathbf{a}^{L+1}_{\mathcal{C}}$ of absolute time of candidate news $c^*$ as $\mathbf{z}^* = [\mathbf{x^*} \mathbf{a}^{L+1}_{\mathcal{C}}] \in \mathbb{R}^{4d}$.

Given the time-aware candidate news representation $\mathbf{z^*}$ and the time-aware news sequence representation $\mathbf{z}_{\mathcal{C}}$, concatenate them together and then perform a linear transformation to transform them into $d$-dimensional space.
Specifically, take news $c_i$ for example, the time-aware representation $\mathbf{z}_{c_i}$ of news $c_i$ is concatenated with $\mathbf{z^*}$ and user embedding $\mathbf{u}$ and then linearly transformed as 
\begin{equation}
    \mathbf{t}_{c_i} = [\mathbf{z}_{c_i} \mathbf{z^*} \mathbf{u}] \mathbf{W_c} \in \mathbb{R}^{d},
\end{equation}
where $\mathbf{W_c} \in \mathbb{R}^{3d \times d}$. With this operation, the sequence-level attention layer can consider content relevance, user preferences, and time dynamics simultaneously. 
The original Transformer is designed for the NLP applications, where the sequential signals are captured by the word indexes.
However, continuous time information can be essential in user behavior modeling, as mentioned before.
Thus, we revise the traditional Transformer by replacing the position encoding with a continuous time embedding.
The variables are firstly input into the self-attention layer,  and then a position-wise feed-forward layer is leveraged to process the output. At last, we use a fully connected layer and a residual connection layer to predict the final results:
\begin{equation} \label{eq:news_vec_1}
\begin{aligned}
        \mathbf{t}_{c_i} &= \text{Attention}(\mathbf{t}_{c_i}), \\
        \mathbf{x}_{c_i} &= \phi{(\mathbf{t}_{c_i} \cdot \mathbf{W_a})}, \\
        \mathbf{x}_{c_i} &= \text{LN}(\text{Dropout}(\mathbf{x}_{c_i} \cdot \mathbf{W_b}) + \mathbf{t}_{c_i}).
\end{aligned}
\end{equation}

where $\phi$ is the activation function, $\mathbf{W_a} \in \mathbb{R}^{d \times d'}$, $\mathbf{W_b} \in \mathbb{R}^{d' \times d}$ are model parameters, and $d'$ is the intermediate size of position-wise feed-forward layer.
``Dropout'' is an approach used for alleviating the overfitting problems, ``LN'' represents layer normalization, which normalizes an input vector by its mean and variance for stable training.

\subsection{History Summarization}
To summarize the users' historical behaviors, we leverage Transformer to process the previously interacted news.
Compared with other sequential models like convolutional neural network (CNN) and recurrent neural network (RNN), Transformer directly captures the correlations between any two steps of events in the sequence, which is effective in many other machine learning tasks. In our model, to obtain the representation of news sequence, we stack the representation of $L$ news articles into a feature tensor $E\in \mathbb{R}^{L\times{d}}$. With $E$ as input, we derive the sequence vector $\mathbf{p}$ by two stacked Transformer layers.

\subsection{Prediction Layer}
Give the sequence vector $\mathbf{p}$, the representation $\mathbf{x}^*$ of candidate news $c^*$ and the user embedding $\mathbf{u}$, we concatenateand and then feed them into a fully-connected layer to estimate the click rate of candidate news $c^*$: 
\begin{equation}
    \hat{y}_{*} = \phi([\mathbf{p}\,\mathbf{x}^*\,\mathbf{u}] \mathbf{W}^{(1)} +\mathbf{b}^{(1)})\mathbf{W}^{(2)} +\mathbf{b}^{(2)},
\end{equation}
where $\mathbf{W}^{(1)}\in\mathbb{R}^{5d\times2d}$, $\mathbf{b}^{(1)}\in\mathbb{R}^{2d}$, $\mathbf{W}^{(2)}\in\mathbb{R}^{2d\times 1}$ and $\mathbf{b}^{(2)}\in\mathbb{R}$ are the parameters of the prediction layer.
Besides, the negative samples set output by dynamic negative sampling layer are combined with positive samples to compute loss. We adopt binary cross-entropy loss function:
\begin{equation}
    L = \sum_{c^* \in D^+ \cup S} {{y}_{*}}\log\sigma({\hat{y}_{*}})+ (1-{y}_{*})\log(1-\sigma(\hat{y}_{*})),
\end{equation}
where $y_{*}$ is the training label of $c^*$, $D^{+}$ represents the positive instance set.

\subsection{Dynamic Negative Sampling}
To train the model, both positive and negative labels are needed. Previous news recommendation models primarily leverage uniform negative sampling to optimize the users' implicit feedback. However, as discussed by \citet{bamler2020extreme}, uniform negative sampling is not optimal since it selects too random samples, which can be easily to be distinguished from the positive ones. To select more discriminative negative samples, we propose a dynamic negative sampling (DNS) method.
In our method, if an training pair is hard to distinguish, we regard it to be more informative and discriminative since the model can learn more by optimizing it. To quatify the hardness of training pair, we introduce a similarity function, that is: $f(\mathbf{y}, \mathbf{X}) = \mathbf{W}(\mathbf{X}^T\cdot \mathbf{y}) + \mathbf{b}$,
where $\mathbf{y}\in \mathbb{R}^d$ and $\mathbf{{X}}\in \mathbb{R}^{d\times N}$ are the representations of the positive and the whole candidate news, respectively. 
$\mathbf{W}\in \mathbb{R}^{N\times N}$ and $\mathbf{d}\in \mathbb{R}^{N}$ are weighting parameters. 
Based on $f$, we select the most similar items in a greedy way: $S = \arg\max f(\mathbf{y}, \mathbf{X})$,
where $S$ is the index set of the selected negative items.

% In our model, we further introduce a loss to constraint the parameters in $f$:
% \begin{equation} 
% L_{S} = \log{(y^T \text{Merge}(S, \mathbf{X}))},
% \end{equation}
% where ``Merge'' is a function projecting the embeddings of the selected negative samples into a vector. We would like to make the negative samples similar to the positive ones on different distance metrics by this loss function.

\section{Experimental Study}\label{sec:exp}
\subsection{Experimental Setup}\label{experimental_setup}
% \noindent
\textbf{Datasets.}
We conduct experiments on three real-world news benchmark datasets: Adressa, Cert and Caing. All these news datasets consist of news articles, anonymous users and their historical clicked news articles with timestamps.
\begin{itemize}
    \item \textbf{Adressa}. This dataset, first introduced by \citet{DBLP:conf/webi/GullaZLOS17}, has been widely used for testing news recommendation systems \citet{10.1145/3543507.3583383, liu2023perconet, 10.1145/3511808.3557335, 10.1145/3383313.3418477, 10.1145/3340531.3411932, zhang2022metonr, wu2023personalized}. Comprising ten weeks of reading logs from Adresseavisen, a Norwegian news portal, the dataset was collected between 1 January to 31 March 2017. In this study, we narrowed our focus to the initial 15 days of collected data for conducting our experiments. The Adressa Dataset, constructed in collaboration between the Norwegian University of Science and Technology (NTNU) and Adressavisen, serves as a rich source of Norwegian news articles across multiple users. 

    \item\textbf{Cert}. The dataset is supplied by the Computer Emergency Response Technical Team of China. It spans from March 2016 to April 2017, encompassing user logs from various Chinese news portals including people.com and cctv.com, among others.
    
    \item \textbf{Caing}. The Caing dataset originates from Caing, a renowned news portal in China (\url{http://www.dcjingsai.com/}). It comprises complete reading logs of 10,000 users over the period of March 2016.
\end{itemize}

Although the MIND dataset \citet{wu2020mind} is currently popular for news recommendation \citet{10.1145/3568022,10.1145/3530257,10.1145/3582425,10.1145/3543507.3583507,10.1145/3568954}, we have excluded it from this study. The dataset is often modeled as an NLP task and notably lacks click time information. Specifically, each news article in the MIND dataset includes a news ID, a title, an abstract, a body, entities, and a category label. The format of each sample is [uID, t, ClickHist, ImpLog], where uID is the anonymous ID of a user, and t is the timestamp of when this sample is constructed. ClickHist consists of an ID list of the news articles previously clicked by this user (sorted by click time \textit{without} clicked timestamp). ImpLog contains the IDs of the news articles and the labels indicating whether they were clicked, with 1 denoting a click and 0 a non-click.
All three datasets are preprocessed to ensure that all users have a minimum of 15 interactions. A summary of the statistical details of these datasets is provided in Table \ref{dataset}.

%%%%%%%%%%%%%%%%%%%%%%%%%%%%%%%%%%%%%%%%
\begin{table}[t]
\centering
\begin{tabular}{llll}
\hline
Datasets& $\#$Interaction& $\#$User&     $\#$News\\
\hline
{Adressa}   & 1,604,879       & 66,649    &   12,034\\
{Cert}      & 1,573,959       & 199       &   588,907\\
{Caing}     & 61,615          & 1,947     &   5,275\\
\hline
\end{tabular}
\caption{{Statistics of the datasets. }}
\label{dataset}
\end{table}
%%%%%%%%%%%%%%%%%%%%%%%%%%%%%%%%%%%%%%%%

% \noindent
\textbf{Baselines.} In our study, we compare {\ourmodel} with several other established methods, which can be broadly categorized into three groups.

The first group comprises collaborative filtering methods:
\begin{itemize}
    \item \textbf{BPR} \citet{DBLP:conf/uai/RendleFGS09}. This method optimizes the latent factor model with the pairwise ranking loss on implicit feedback data.

    \item \textbf{GRU4Rec} \citet{DBLP:journals/corr/HidasiKBT15}. This approach uses a recurrent neural network for session-based recommendations.

    \item \textbf{Caser} \citet{DBLP:conf/wsdm/TangW18}. This method utilizes horizontal and vertical convolution filters to simultaneously capture sequential patterns and model users' general preferences.
\end{itemize}

The second group includes recommendation methods that utilize content information:
% \noindent
% $\bullet$ \textbf{GRU4Rec++} \citet{hidasi2018recurrent}:
% This method is an improved version of GRU4Rec, further considering news content information.

% \noindent
% $\bullet$ \textbf{WE3CN}\citet{DBLP:conf/cikm/KhattarKV018a}:
% This method applies 3D convolution neural network for news recommendation, utilizing content and sequential information simultaneously.

% \noindent
% $\bullet$ \textbf{NRMS} \citet{wu2019neural}:

% \noindent
% $\bullet$ \textbf{NPA} \citet{wu2019npa}:

% \noindent
% $\bullet$ \textbf{LSTUR} \citet{an2019neural}:

\begin{itemize}
    \item \textbf{NAML} \citet{wu2019neural2}. It is a neural news recommendation approach, uses different types of news information to learn informative news and user representations with the help of a news encoder and a user encoder.

    \item \textbf{TEKGR} \citet{10.1145/3340531.3411932}. This model suggests a knowledge graph-enhanced recommendation system where knowledge is extracted from a knowledge graph based on news articles' contextual information. The extracted knowledge is then used to enrich the news representation.

    \item \textbf{WG4Rec} \citet{10.1145/3459637.3482401}. This model proposes a new textual content representation method by creating a word graph for recommendation, using three types of word associations (including semantically-similar word vectors, co-occurrence in documents and co-click by users) for content representation and user preference modeling.

    % \item \textbf{ATNR} \citet{ZHU202233}. This approach uses three encoders for news, users, and sequences, respectively, learning the deep interactions between users and candidate news.

    \item \textbf{SI-News} \citet{ZHU202233}. This model employs an attentional graph convolutional network (GCN) to embed users’ interests from their social information and enhances news representations by optimizing the titles and contents of news via attention mechanisms.

    % \item \textbf{MFF} \citet{10.1145/3555373}. This model proposes a multi-factor fusion approach for news recommendation, integrating user-dependent preference effect and user-independent timeliness effect.
\end{itemize}

The third group contains recommendation methods that utilize dynamic information:
\begin{itemize}
    \item \textbf{DNA} \citet{zhang2019dynamic}: This approach uses a hierarchical attention network made up of convolution neural networks to model sentence-, element-, and news-level information, using a manually designed exponential decay factor to indicate click order instead of the ground-truth clicked time.

    % \item \textbf{Time-LSTM} \citet{DBLP:conf/ijcai/ZhuLLWGLC17}: This method underscores the importance of time intervals between users’ actions for capturing their relations and proposes a LSTM variant with time gates to model these intervals.

    \item \textbf{TiSASRec} \citet{li2020time}: This model addresses the gap in most models that only regard interaction histories as ordered sequences without considering the time intervals between each interaction. TiSASRec seeks to model the timestamps of interactions within a sequential modeling framework, exploring the influence of different time intervals on next item prediction. It proposes to use both the absolute positions of items and the time intervals between them in a sequence.
\end{itemize}

% \noindent
\textbf{Evaluation Protocols.}
As for performance evaluation, we adhere to the protocol established by \citet{li2020time,10.1145/3459637.3482401,HUANG2022119,10.1145/3555373}, using the Hit Ratio (HR) and Normalized Discounted Cumulative Gain (NDCG). HR@N measures if the ground-truth news item is ranked within the top-N list, while NDCG@N takes into account the hit's position, awarding higher scores to hits at the top ranks.

% \noindent
\textbf{Implementation.}
To extract news elements, we utilize Polyglot \citet{al2015polyglot} for the Norwegian Adressa dataset, given its support for multilingual applications. For the Chinese datasets Cert and Caing, we employ NLPIR \url{http://ictclas.nlpir.org} (also known as ICTCLAS) given its specialized capabilities for the Chinese language. 
Given the time-intensive nature of ranking all news items, we opt for sampling unclicked news as negative labels. Specifically, during the training phase, we apply a dynamic negative sampling strategy to generate more discriminative samples. In the inference phase, however, we randomly select negative samples to ensure a fair comparison with existing methods. In our experiments, we balance the candidate news with 99 negative news samples and 1 positive news sample (the clicked one). We employ a sliding window of length $L+1$ ($L$ historical news items and $1$ candidate news item) to traverse through their interactions, generating one instance per window. We reserve the most recent instance for testing and use the rest for training.
In terms of hyperparameters, for the baseline models BPR, Caser, and GRU4Rec, we set both user and news embedding dimensions at 20. For GRU4Rec++, we set the content vector dimension at 64. We adopt default settings for the other parameters in the baseline models. The number of news items $L$ in each sequence is set to 10, while the number of sentences $K$ for each news item is set to 20. We set the dimension $d$ at 64 and the intermediate dimension size $d'$ at 256.
We employ the Adam optimizer for training, with the learning rate, batch size, weight decay, and dropout set to $10^{-3}$, 256, $10^{-4}$, and 0.2, respectively. To ensure consistency, each experiment is replicated three times, and we report the average results.

\textbf{Research Questions.} Through our experiments, we aim to answer the following pivotal questions:

\noindent
$\bullet$ \textbf{RQ1}: Can \ourmodel~surpass the performance of several recent news recommendation models?

\noindent
$\bullet$ \textbf{RQ2}: Do absolute and relative time contribute to improving model performance?

\noindent
$\bullet$ \textbf{RQ3}: Does dynamic negative sampling  enhance recommendation performance?

\noindent
$\bullet$ \textbf{RQ4}: What impact does each layer in the hierarchy have on the overall performance?

\noindent
$\bullet$ \textbf{RQ5}: Are the attention weights effective in learning meaningful information at the sentence-, element-, and sequence-levels?

% \noindent
% $\bullet$ \textbf{RQ6}: Can the history summarization layer further improve the performance of hierarchical attention layers?

We detail our findings related to these research questions in the following sections. Note that the answers to these questions shed light on the effectiveness and utility of \ourmodel~in the field of news recommendation.

\subsection{Performance Comparison (RQ1)}\label{RQ1}
% %%%%%%%%%%%%%%%%%%%%%%%%%%%%%%%%%%%
\begin{table*}[!t]
\centering
\begin{tabular}{lcccccc}
\toprule
\textbf{Adressa}&   HR@1&   HR@5&   HR@10&   NDCG@1&   NDCG@5&   NDCG@10 \\
\hline
BPR&        0.0777&    0.2676&    0.4278&    0.0777&    0.1724&    0.2239 \\
GRU4Rec&    0.2969&    0.6926&    0.8535&    0.2969&    0.5026&    0.5551 \\
Caser&      0.3327&    0.7277&    0.8684&    0.3327&    0.5397&    0.5856 \\
\cdashline{1-7}[0.8pt/2pt]
NAML&       0.4903&    0.8910&    0.9521&    0.4990&    0.6833&    0.7087\\
TEKGR&      0.5391&    0.8980&    0.9534&    0.5210&    0.6951&    0.7231\\
WG4Rec&     0.5689&    0.8979&    0.9550&    0.5763&    0.7259&    0.7408\\
Si-News&    \underline{0.5976}&    \underline{0.8997}&    \underline{0.9648}&    \underline{0.6045}&    \underline{0.7523}&    \underline{0.7763}\\
\cdashline{1-7}[0.8pt/2pt]
DNA&        0.4528&    0.8627&    0.9505&    0.4528&    0.6726&    0.7015 \\
TiSASRec&   0.5788&    0.8976&    0.9576&    0.5832&    0.7301&    0.7577\\
{\ourmodel}&\textbf{0.6355}&    \textbf{0.9324}&    \textbf{0.9761}&    \textbf{0.6355}&    \textbf{0.7977}&    \textbf{0.8119} \\
\bottomrule

\textbf{Cert}&   HR@1   &   HR@5   &   HR@10   &   NDCG@1   &   NDCG@5   &   NDCG@10 \\
\hline
BPR&        0.2463&    0.4539&    0.5444&    0.2463&    0.3511&    0.3801\\
GRU4Rec&    0.3367&    0.4623&    0.5193&    0.3367&    0.4038&    0.4222\\
Caser  &    0.4556&    0.5812&    0.6281&    0.4556&    0.5251&    0.5403\\
\cdashline{1-7}[0.8pt/2pt]
NAML&       0.5539&    0.8286 &   0.9071 &   0.5333&    0.6936&    0.7187\\
TEKGR&      0.6719&    0.8646&    0.9293&    0.6649&    0.7620&    0.8093\\
WG4Rec&     0.7216&    0.9037&    0.9500&    0.7121&    0.8016&    0.8383\\
Si-News&    \underline{0.7512}&    \underline{0.9277}&    \underline{0.9749}&    \underline{0.7502}&    \underline{0.8362}&    \underline{0.8506}\\
\cdashline{1-7}[0.8pt/2pt]
DNA&        0.5239&    0.8116&    0.8920&    0.5239&    0.6746&    0.7007\\
TiSASRec&   0.7135&    0.9004&    0.9460&    0.7026&    0.7925&    0.8356\\
{\ourmodel}&\textbf{0.7513}&    \textbf{0.9346}&    \textbf{0.9774}&    \textbf{0.7513}&    \textbf{0.8380}&    \textbf{0.8515}\\
\bottomrule

\textbf{Caing}&   HR@1   &   HR@5   &   HR@10   &   NDCG@1   &   NDCG@5   &   NDCG@10 \\
\hline
BPR&        0.3546&    0.5728&    0.6774&    0.3546&    0.4699&    0.5038\\
GRU4Rec&    0.4633&    0.7713&    0.8541&    0.4633&    0.6314&    0.6586\\
Caser&      0.5999&    0.7964&    0.8514&    0.5999&    0.7057&    0.7235\\
\cdashline{1-7}[0.8pt/2pt]
NAML&       0.6330&    0.8408&    0.9043&    0.6394&    0.7481&    0.7696\\
TEKGR&      0.6498&    0.8487&    0.9092&    0.6431&    0.7521&    0.7747\\
WG4Rec&     0.6756&    0.8596&    0.8929&    0.6658&    0.7579&    0.7702\\
Si-News&    \underline{0.7092}&    \underline{0.8820}&    \underline{0.9190}&    \underline{0.7046}&    \underline{0.7921}&    \underline{0.8098}\\
\cdashline{1-7}[0.8pt/2pt]
DNA&        0.6220&    0.8391&    0.9033&    0.6220&    0.7401&    0.7609\\
TiSASRec&   0.6691&    0.8437&    0.8882&    0.6560&    0.7429&    0.7667 \\
{\ourmodel}&\textbf{0.7244}&    \textbf{0.8942}&    \textbf{0.9361}&    \textbf{0.7244}&    \textbf{0.8153}&    \textbf{0.8294}\\
\bottomrule
\end{tabular}
\caption{Performance comparison on three datasets for all methods. }
\label{performance}
\end{table*}
%%%%%%%%%%%%%%%%%%%%%%%%%%%%
In this section, we evaluate the effectiveness of our proposed method, referred to as {\ourmodel}, against multiple baseline methods for news recommendation using three different datasets: Adressa, Cert, and Caing. The comparison focuses on six evaluation metrics, namely Hit Rate (HR) and Normalized Discounted Cumulative Gain (NDCG) at cut-offs 1, 5, and 10.

In the Adressa dataset, {\ourmodel} performs better in all metrics. It surpasses the second-best method, Si-News, by around (3.79\%, 3.27\%, 1.13\%, 3.10\%, 4.54\%, 3.56\%) in (HR@1, HR@5, HR@10, NDCG@1, NDCG@5, NDCG@10), respectively. 
On the Cert dataset, {\ourmodel} again showcases strong performance. It equals the best HR@1 score of 0.7513 achieved by Si-News and surpasses all other methods on HR@5 and HR@10, demonstrating its proficiency in recommending relevant news articles even at a larger cut-off. In terms of NDCG@1 and NDCG@5, {\ourmodel} exceeds the closest competitor, Si-News, by 0.11 and 0.18 points, respectively. However, for NDCG@10, it falls slightly behind Si-News, hinting at potential areas for future improvement.
For the Caing dataset, {\ourmodel} also leads in HR@1, HR@5, HR@10, NDCG@1, and NDCG@5, outperforming the runner-up, Si-News, by about 1.52\% in HR@1, 1.22\% in HR@5, 1.71\% in HR@10, 1.98\% in NDCG@1, and 2.32\% in NDCG@5. However, just like in the Cert dataset, it has a slightly lower NDCG@10 score compared to Si-News.
In conclusion, our method, {\ourmodel}, consistently shows superior performance over the baselines across all three datasets in most of the metrics, thereby confirming its effectiveness in the news recommendation task. The areas where it doesn't lead could potentially be improved in future work, making it even more robust and accurate.

\subsection{Influence of Time Embedding (RQ2)}

%%%%%%%%%%%%%%%%%%%%%%%%%%%
\begin{figure}[t]
\centering
\includegraphics[width=0.6\columnwidth]{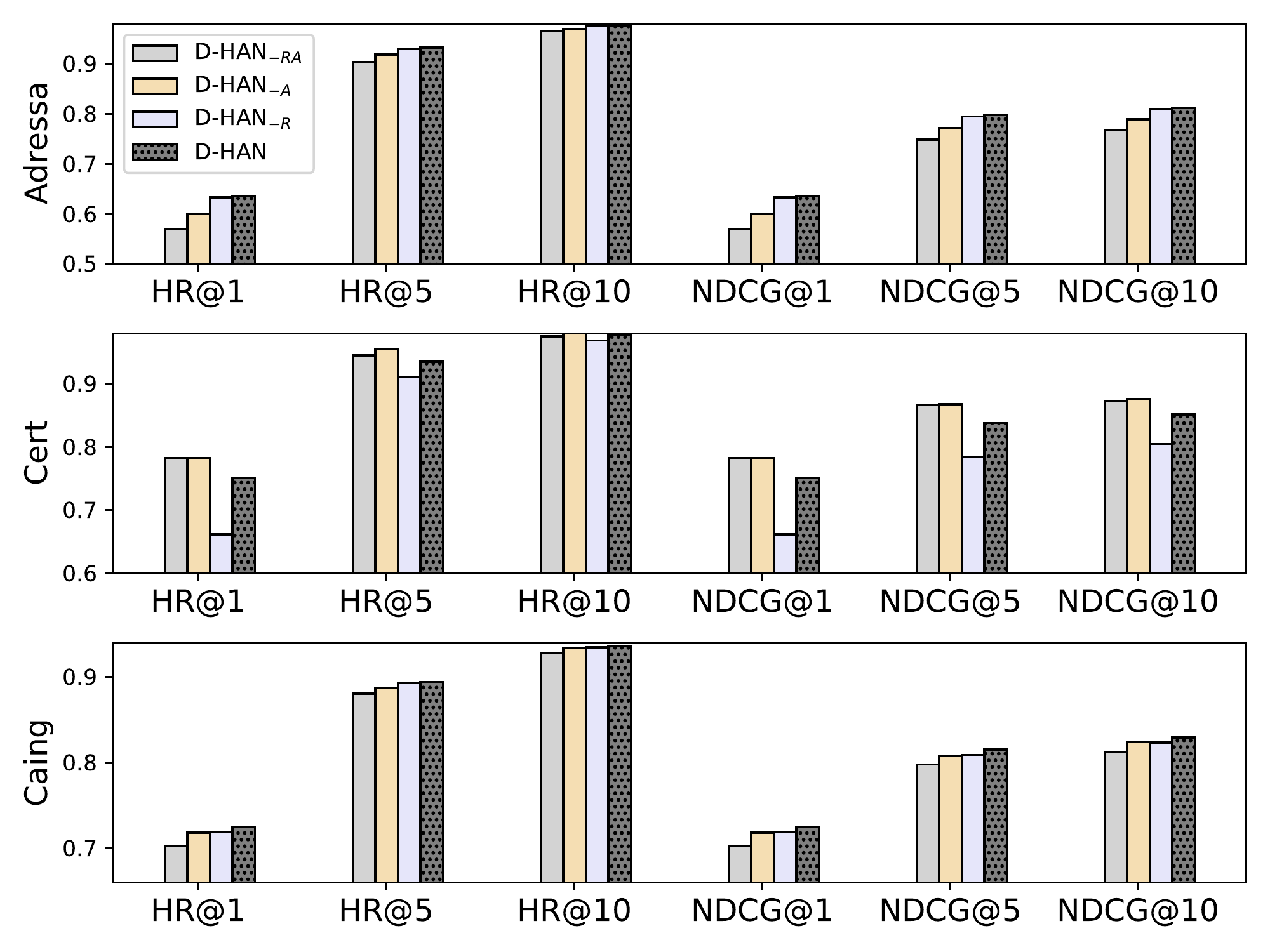}
\caption{Ablation study on dynamics information integration.}
\label{dhan_ablation}
\end{figure}
%%%%%%%%%%%%%%%%%%%%%%%%%%%
The ablation study investigates the performance of different versions of \ourmodel~by selectively removing time-related components, as shown in Figure~\ref{dhan_ablation}. 
In the Adressa dataset, \ourmodel~outperforms all of its component-removed variants in all metrics, demonstrating the contribution of each component to the final performance. This underlines the effectiveness of the complete \ourmodel~model. Interestingly, removing relative time ({\ourmodel}$_{-\text{R}}$) seems to have the least impact on the performance of the model across all metrics, whereas removing absolute time ({\ourmodel}$_{-\text{A}}$) or both relative and absolute time ({\ourmodel}$_{-\text{RA}}$) results in more substantial performance drop.
In the Cert dataset, however, the situation is slightly different. While \ourmodel~performs better than the variant where relative time is removed, it does not surpass the variant without absolute time in all metrics. This might suggest that for this particular dataset, the absolute timing information contributes less to the model's performance.
For the Caing dataset, we again observe that \ourmodel~outperforms all of its variants across all metrics. This dataset follows a similar pattern to the Adressa dataset, demonstrating that the full feature set of \ourmodel~generally leads to the best performance.
This implies that these temporal features might interact with dataset-specific characteristics.

\subsection{Impact of Dynamic Negative Sampling (RQ3)}
In this subsection, we examine the impacts of the Dynamic Negative Sampling (DNS). To ensure a fair comparison, we incorporate DNS during the training phase while opting for uniform negative sampling during the testing phase. Owing to space constraints, we focus on the results for HR@1, NDCG@5, and NDCG@10, as shown in Table \ref{negative_sampling}. It can be observed that DNS generally enhances model performance. These findings corroborate the efficacy of our proposed dynamic negative sampling method. In comparison with uniform sampling, DNS is capable of identifying more informative negative samples that have a higher discriminative ability.
% ======================================================
\begin{table}[t]
\centering
\begin{tabular}{cccc}
\toprule
\textbf{Adressa}&   HR@1&  NDCG@5&   NDCG@10 \\
\midrule
{\ourmodel}$_U$&   \textbf{0.5685}&    0.7486&    0.7676  \\
{\ourmodel}$_D$&   0.5666&    \textbf{0.7508}&    \textbf{0.7698} \\
\midrule
\midrule
\textbf{Cert}&   HR@1&  NDCG@5&   NDCG@10 \\
\midrule
{\ourmodel}$_U$&   0.7822&    0.8662&    0.8726  \\
{\ourmodel}$_D$ &   \textbf{0.7915}&    \textbf{0.8782}&    \textbf{0.8856} \\
\midrule
\midrule
\textbf{Caing}&   HR@1&  NDCG@5&   NDCG@10 \\
\midrule
{\ourmodel}$_U$&   0.7026&    0.7979&    0.8119  \\
{\ourmodel}$_D$&   \textbf{0.7089}&    \textbf{0.8031}&    \textbf{0.8182} \\
\bottomrule
\end{tabular}
\caption{Comparison of different negative sampling methods. {\ourmodel}$_U$ refers to uniform negative sampling is used, while {\ourmodel}$_D$ denotes dynamic negative sampling is used. In this report, dynamics information, i.e., time information, is omitted.}
\label{negative_sampling}
\end{table}
% ======================================================

\subsection{Impacts of hierarchies (RQ4)}
\begin{figure}[h]
\centering
\includegraphics[width=0.6\columnwidth]{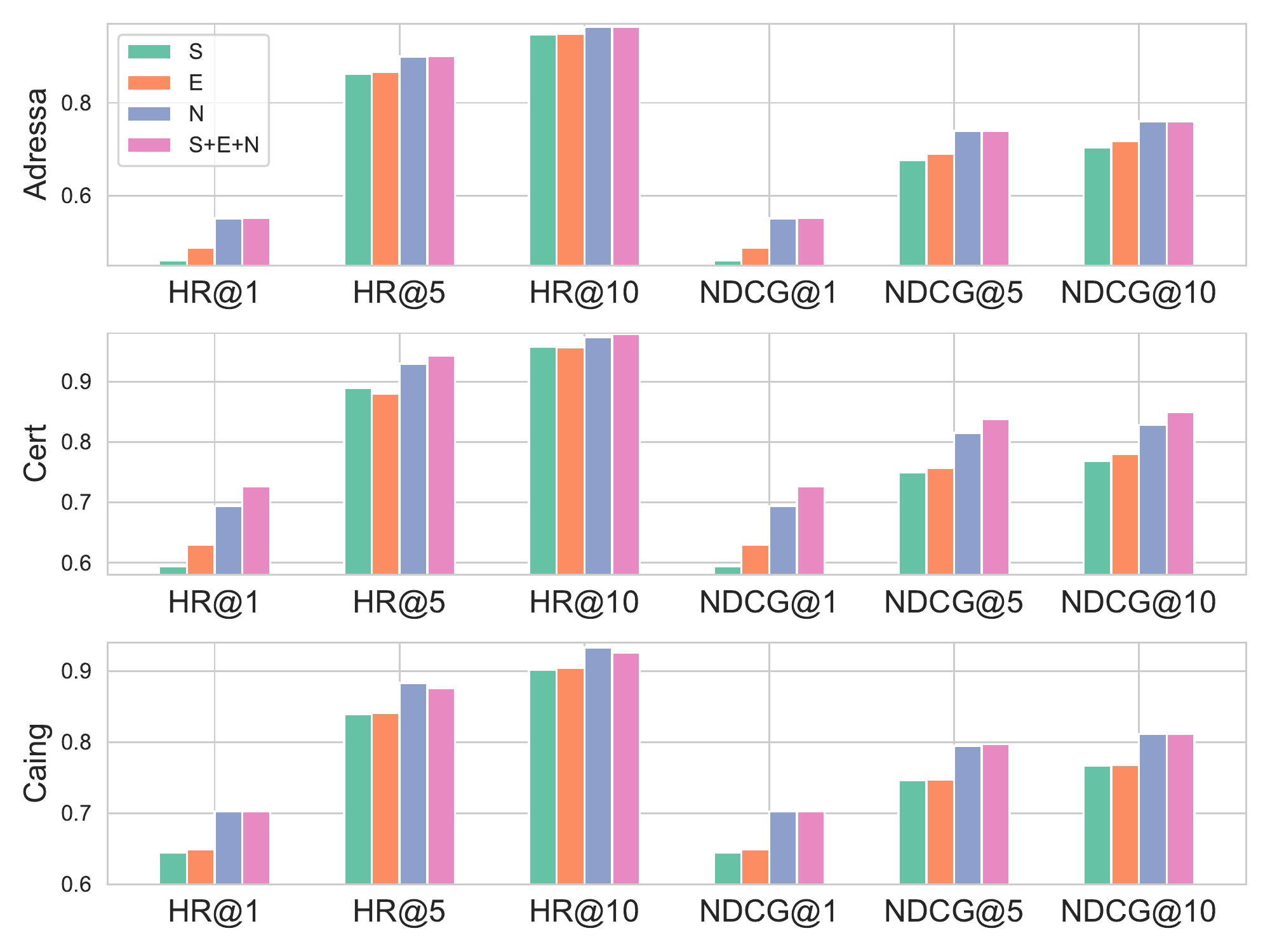}
\caption{Ablation study on hierarchy component. S, E, N represents sentence-, element-, sequence-level component is solely used, respectively. S+E+N refers to they are used in combination.}
\label{hierarch_ablation}
\end{figure}

In this section, we make an ablation study about the impacts of each hierarchy, as shown in Figure~\ref{hierarch_ablation}.
Looking at the results for the Adressa dataset, we observe that our model with sentence- (S), element- (E), and sequence-level (N) attention components combined (S+E+N) consistently outperforms the other configurations across all metrics. Moreover, for the role of each hierarchy, sequence-level components have more significant impacts on the overall performance. The Cert dataset results show a similar trend. The combination of all three layers (S+E+N) achieves the highest performance across all the metrics. This demonstrates the effectiveness of incorporating various attention mechanisms at different granularities in our model. When considering the Caing dataset, we see a similar pattern with S+E+N again performing the best in HR@1, NDCG@1, NDCG@5, and NDCG@10. However, for the HR@5 and HR@10 metrics, the model with only the sequence-level attention layer (N) performs slightly better than the combined approach. In conclusion, the data indicates that a model that leverages attention at multiple granularities simultaneously (sentence-, element-, and sequence-levels) generally outperforms models that use these mechanisms individually. This suggests the advantage of a multi-faceted approach that captures dependencies and patterns across different levels of the data hierarchy. 

%%%%%%%%%%%%%%%%%%%%%%%%%%%
%%%%%%%%%%%%%%%%%%%%%%%%%%%

\subsection{Visualization of Attention Weights (RQ5)}
\begin{figure*}[t]
\centering
\subfigure[Sentence-level ${w}$]{
\begin{minipage}[t]{0.23\linewidth}
\centering
\includegraphics[width=2in]{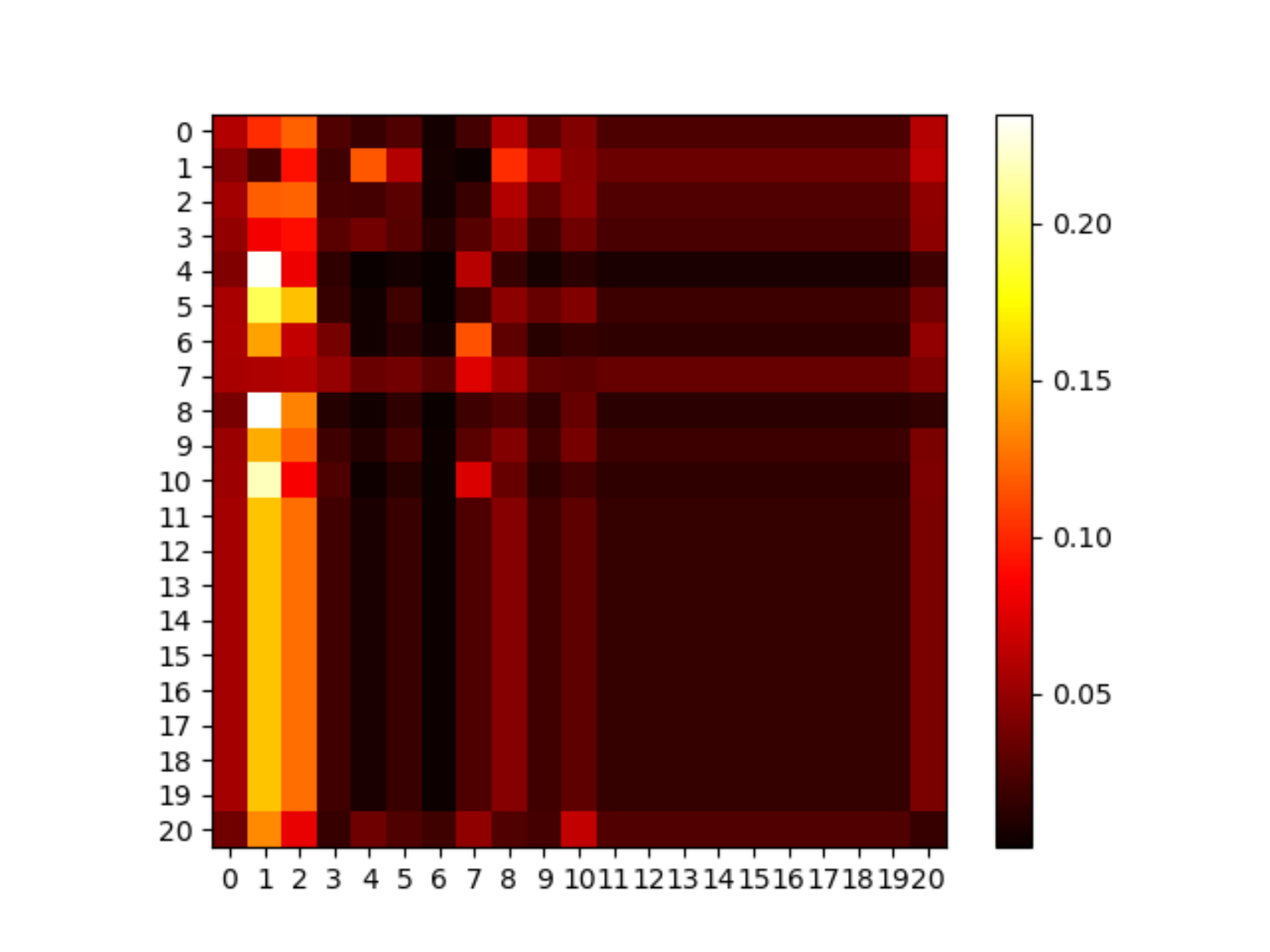}
\end{minipage}%
\label{fig_a}
}%
\subfigure[Element-level ${w}$]{
\begin{minipage}[t]{0.23\linewidth}
\centering
\includegraphics[width=2in]{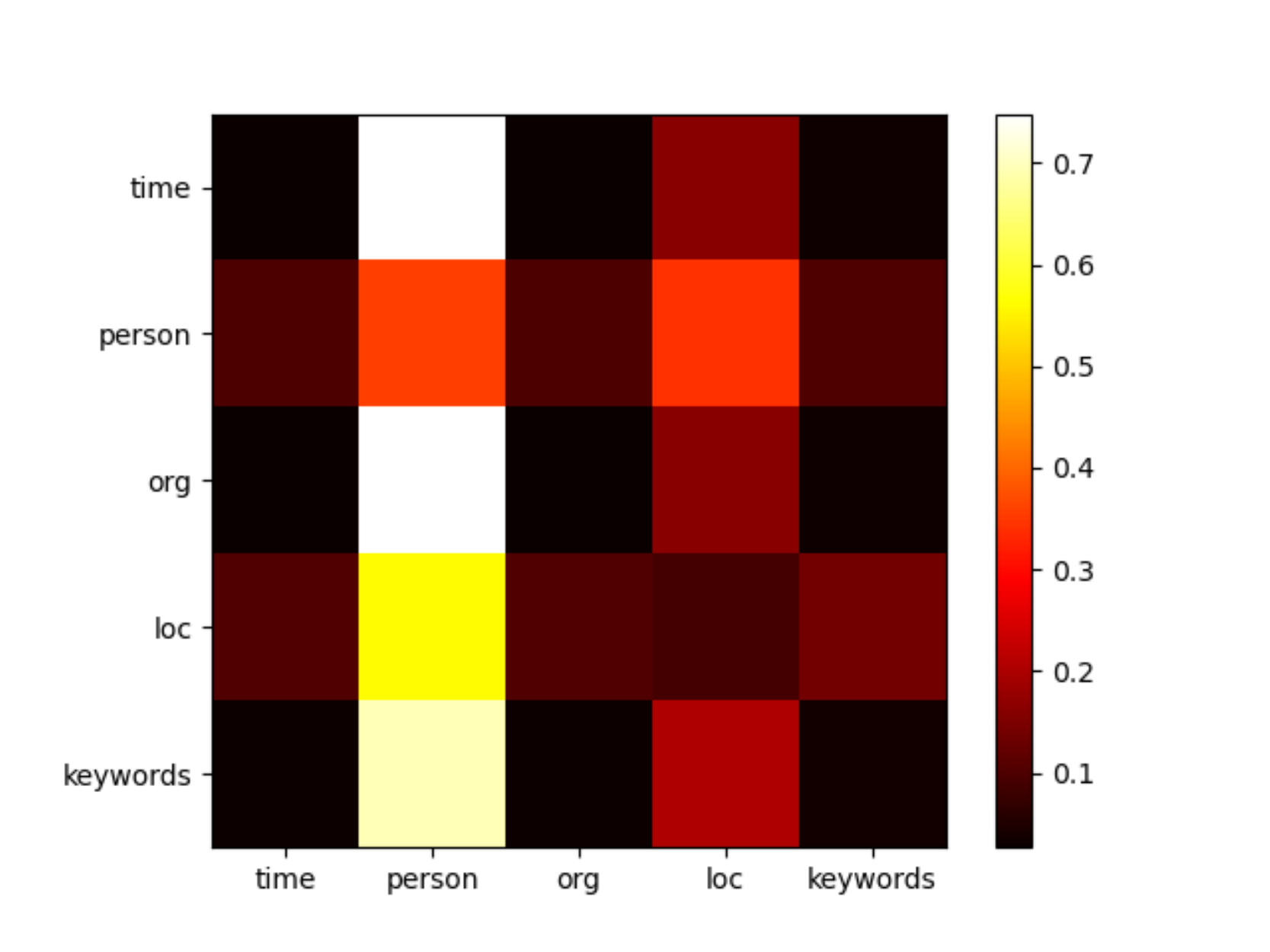}
\end{minipage}%
\label{fig_b}
}%
\subfigure[Sequence-level ${w}$]{
\begin{minipage}[t]{0.23\linewidth}
\centering
\includegraphics[width=2in]{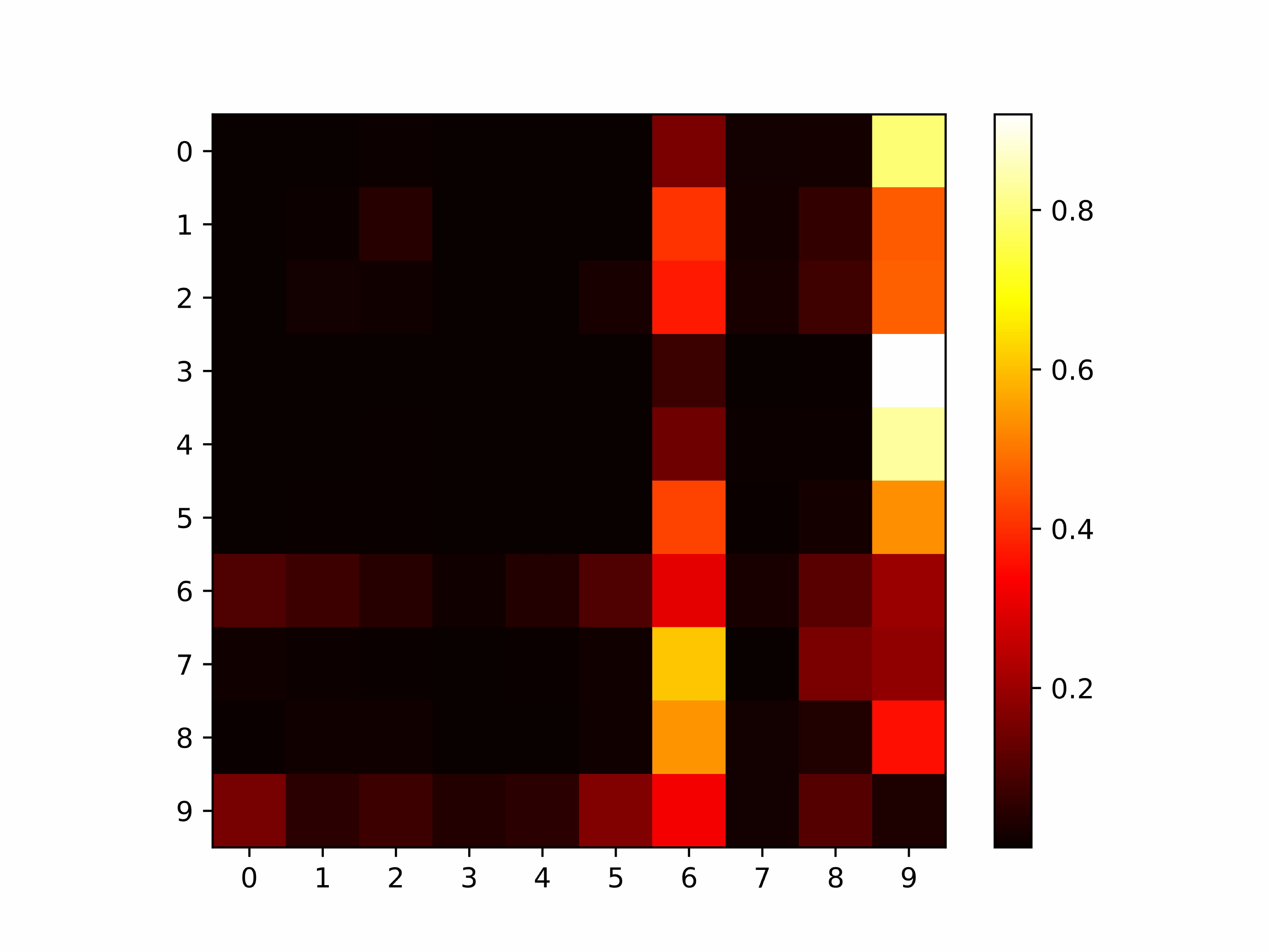}
\end{minipage}
\label{fig_c}
}%
\subfigure[Time-aware sequence-level ${w}$]{
\begin{minipage}[t]{0.23\linewidth}
\centering
\includegraphics[width=2in]{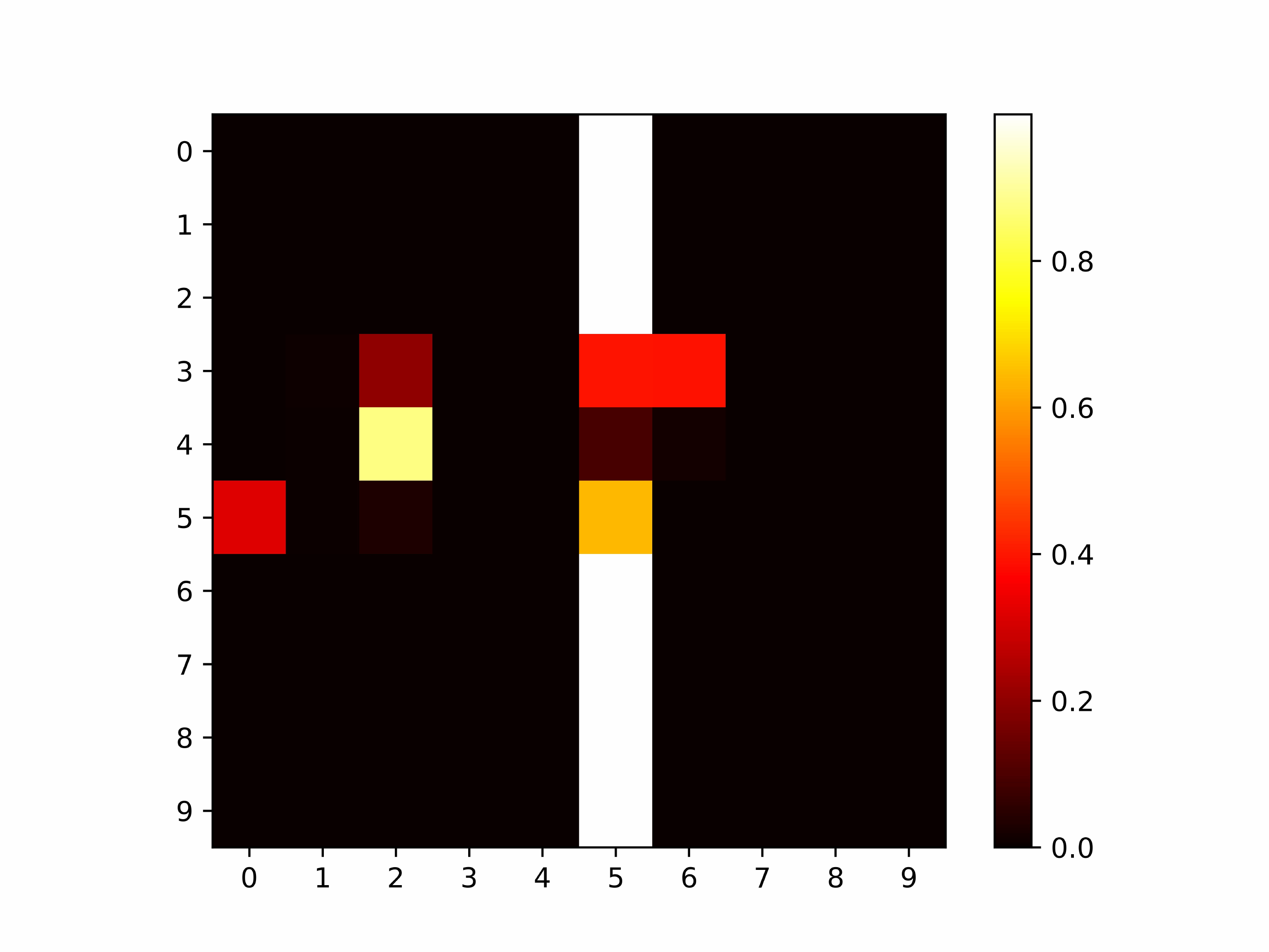}
\end{minipage}
\label{fig_d}
}%
\centering
\caption{Visualization of sentence-, element-, sequence- and time-aware sequence-level attention weights between historical news sequence and candidate news. ${w}$ represents attention weights.}
\label{case_study_1}
\end{figure*}

In this subsection, we present visual case studies of the sentence-, element-, sequence- and time-aware sequence-level attention weights to demonstrate what each module of our model learns. We randomly sample pairs of historical and candidate news from the testing set, and the attention weights are extracted from the epoch with the best performance. For a fair comparison, we use the same instance for figures \ref{fig_c} and \ref{fig_d}. Due to space limitations, we only provide visualizations from the Caing dataset.

\textbf{Sentence-level attention weights.}
Figure \ref{fig_a} presents the self-attention weights of a history-candidate news pair, wherein the candidate news representation is appended as the last sentence of the historical news, as detailed in Section \ref{sen_attn_model}. The visualization shows that sentences tend to focus on the first few sentences, with less attention paid to sentences in the latter part. Notably, the first three sentences receive significant attention, in line with the text characteristics in the news domain where the main content of the news is often contained within the initial few sentences. Moreover, the last column and row have higher attention weights than their adjacent ones, but the bottom right corner (candidate news attention weight) is lower. This could be because the candidate news provides reference information for historical news, but not for itself.

\textbf{Element-level attention weights.}
Figure \ref{fig_b} presents a heatmap of attention weights between two elements of a history-candidate news pair, with this pair concatenated along the $d$ dimension as described in Section \ref{ele_attn_model}. The x-axis and y-axis represent the five elements of news we use, i.e., time, person, organization, location, and keywords. The visualization indicates that the person and location elements play significant roles, implying their importance in this instance. Unlike the fixed and typical patterns in the heatmap of sentence-level attention weights, the heatmap of element-level attention weights varies across different instances. This may be because the five selected elements are already essential for news sequence recommendation.

\textbf{sequence-level attention weights.}
Figure \ref{fig_c} shows a typical heatmap of sequence-level attention weights between news items in a news sequence. It's clear that the upper left part of the figure has lower weights, while the bottom, left, and bottom left have higher weights. Specifically, the 9$^{th}$ and 6$^{th}$ news items have the highest weights, indicating their influence on user decisions in this instance. Furthermore, we sampled 500 instances from the test set and categorized the influence of distant and recent news into three categories: (a) recent news have larger influence, (b) distant news have larger influence, (c) older news have the same influence as recent news. The distribution of these categories is 46.8\%, 23.6\%, and 29.6\%, respectively, suggesting that user behavior patterns in news recommendation scenes are more affected by recent behavior.

\textbf{Time-aware sequence-level attention weights.}
Given the importance of time intervals and timestamps in user behaviors, we consider how they affect the attention weights. For instance, a user might be highly interested in a newly released news article, but their interest could wane after a certain period. Figure \ref{fig_d} shows the heatmap of sequence-level attention weights of Figure \ref{fig_c}, augmented with time information. Comparing Figures \ref{fig_d} and \ref{fig_c}, we can see that the time-aware sequence-level attention weights are sparser, and weights are dominated by the 5$^{th}$ news item. This suggests that embedding relative time intervals and absolute timestamps significantly influences the attention weights.

%%%%%%%%%%%%%%%%%%%%%%%%%%%%
%%%%%%%%%%%%%%%%%%%%%%%%%%%%%%%

\subsection{Analysis of History Summarization Layer}
This section conducts an analysis on the importance of the history summarization layer. The approach involves testing different sequential models for user history summarization, including CNN, LSTM, BiLSTM, single head Transformer (SH), and multi-head Transformer (MH). The experiment also explores the performance of the model when the history summarization layer is removed, i.e., {\ourmodel}$_{-\text{S}}$. The results in Table \ref{history_sum} reveal the following observations. (a) Including a history summarization layer improves the model's performance across all three datasets. On the Adressa dataset, using CNN, LSTM, or BiLSTM as the history summarization layer enhances the model's performance, with CNN performing best on the Cert dataset. However, on the Cert and Caing datasets, RNN-based models do not improve the model's performance. This indicates that CNN and RNN-based summarization methods may have unique advantages on specific datasets. However, it is worth noting that Transformer-based models consistently deliver the best performance, presumably because the Transformer is capable of modeling the correlations between any two steps in the news sequence, thereby capturing sequential patterns more effectively.
(b) Comparing the results of the single-head (SH) with the multi-head (MH) Transformer shows that having more heads does not necessarily lead to a significant improvement in performance. In our experiments, we opted for a single-head Transformer for simplicity.

%%%%%%%%%%%%%%%%%%%%%%%%%%%%%%%%%%%%%%%%%%
\begin{table*}[!htp]
\centering
\begin{tabular}{lcccccc}
\toprule
\textbf{Adressa}&   HR@1&   HR@5&   HR@10&   NDCG@1&   NDCG@5&   NDCG@10 \\
\midrule
{\ourmodel}$_{-\text{S}}$  &   0.5518&    0.9001&  0.9630& 0.5518&  0.7398&  0.7604    \\
\hdashline
CNN     &     0.5551&    0.9014&  0.9638&    0.5551&  0.7419&  0.7620    \\
LSTM    &   0.5593&    0.9036&    0.9653&   0.5593& 0.7456&    0.7659 \\
BiLSTM  &   0.5621&    0.9057&    0.9658&  0.5621&  0.7473&  0.7671  \\
\hdashline
SH      &   \textbf{0.5685}&    0.9031&    0.9653&  \textbf{0.5685}&  0.7486&  0.7676  \\
MH=2    &   0.5653&    \textbf{0.9059}&    \textbf{0.9662}&  0.5653&  \textbf{0.7499}&  \textbf{0.7677} \\
MH=4    &   0.5634&    0.9054&    0.9656&  0.5633&  0.7477&  0.7673 \\
MH=8    &   0.5636&    0.9041&    0.9651&  0.5635&  0.7469&  0.7665 \\
\hline\hline

\textbf{Cert}&    HR@1&   HR@5&   HR@10&   NDCG@1&   NDCG@5&   NDCG@10 \\
\midrule
{\ourmodel}$_{-\text{S}}$  &   0.7270&    0.9430&  0.9782&  0.7270&  0.8381&  0.8499   \\
\hdashline
CNN&   \textbf{0.7889}&    0.9464&  0.9749&  \textbf{0.7889}&  \textbf{0.8687}&  \textbf{0.8791}   \\
LSTM&   0.7236&    0.9112&    0.9698&    0.7236&   0.8172&  0.8280  \\
BiLSTM&   0.7085&    0.9028&    0.9664&    0.7085&   0.8042&  0.8228  \\
\hdashline
SH    &   0.7822&    0.9447&   0.9749&   0.7822&   0.8662&  0.8726 \\
MH=2  &   0.7672&    0.9447&    0.9732&   0.7672&   0.8531&  0.8618 \\
MH=4  &   0.7822&    \textbf{0.9548}&    0.9799&   0.7822&   0.8677&  0.8723   \\
MH=8  &   0.7571&    0.9498&    \textbf{0.9832}&   0.7571&   0.8560&  0.8643   \\
\hline\hline

\textbf{Caing}&    HR@1&   HR@5&   HR@10&   NDCG@1&   NDCG@5&  NDCG@10 \\
\midrule
{\ourmodel}$_{-\text{S}}$&    0.7024&    0.8759&  0.9256&  0.7024&  0.7969&  0.8115    \\
\hdashline
CNN&  0.6968&    0.8752&  0.9276&  0.6968&  0.7925&  0.8093    \\
LSTM&   0.6901&    0.8639&    0.9175&   0.6901&  0.7801&   0.7978  \\
BiLSTM&   0.6976&    0.8725&    0.9245&   0.6976&  0.7902&   0.8067 \\
\hdashline
SH&   0.7026&    0.8804&    0.9279&   0.7026&  \textbf{0.7979}&  0.8119  \\
MH=2&   \textbf{0.7062}&  0.8805&    0.9291&   \textbf{0.7062}&  0.7972&   0.8126 \\
MH=4&   0.7049&    0.8818&    \textbf{0.9309}&  0.7049&  0.7971&  0.8129  \\
MH=8&   0.7054&    \textbf{0.8819}&    0.9305&  0.7054&  0.7971&  \textbf{0.8131}  \\
\bottomrule
\end{tabular}
\caption{Comparison of different implementations of the history summarization layer. {\ourmodel}$_{-\text{S}}$ refers to {\ourmodel} without the history summarization layer. CNN, LSTM, BiLSTM, SH (Single Head), and MH=2 (Multi-Head with 2 heads) represent different implementations of the history summarization layer within {\ourmodel}. The bolded numbers indicate the best result for each column. In this comparison, time information is not included.}
\label{history_sum}
\end{table*}
%%%%%%%%%%%%%%%%%%%%%%%%%%%%%%%%%%%%%%%%%%%%

\section{Conclusion}\label{sec:con}
We introduced a hierarchical attention network for news recommendation that integrates sentence-, element-, and sequence-level information. This approach also utilizes a Transformer encoder to capture the sequential patterns inherent in news interactions. To better model the dynamic nature of news recommendation, we incorporated a time-aware sequence-level attention layer that takes into account both relative time intervals and absolute timestamps.
To further enhance our model's performance, we proposed a dynamic negative sampling method. This method generates negative instances dynamically during training, providing more informative negative samples that can guide the optimization process more effectively.

\justifying 
\section*{Acknowledgments}
This work is greatly enhanced by the insightful feedback and suggestions from the reviewers of this manuscript. Their valuable comments and critiques have substantially improved the quality and clarity of this paper.
We would like to extend our heartfelt appreciation for their contributions.
Lastly, we gratefully acknowledge the support of the National Natural Science Foundation of China (NSFC) under Grant No. 61925203 and China Scholarship Council (CSC) under Grant No. 202206020120. 
\bibliographystyle{elsarticle-num-names}  
\bibliography{D-HAN} % remove under line

\begin{thebibliography}{51}
\expandafter\ifx\csname natexlab\endcsname\relax\def\natexlab#1{#1}\fi
\providecommand{\url}[1]{\texttt{#1}}
\providecommand{\href}[2]{#2}
\providecommand{\path}[1]{#1}
\providecommand{\DOIprefix}{doi:}
\providecommand{\ArXivprefix}{arXiv:}
\providecommand{\URLprefix}{URL: }
\providecommand{\Pubmedprefix}{pmid:}
\providecommand{\doi}[1]{\href{http://dx.doi.org/#1}{\path{#1}}}
\providecommand{\Pubmed}[1]{\href{pmid:#1}{\path{#1}}}
\providecommand{\bibinfo}[2]{#2}
\ifx\xfnm\relax \def\xfnm[#1]{\unskip,\space#1}\fi
%Type = Inproceedings
\bibitem[{Dwivedi and Arya(2016)}]{dwivedi2016survey}
\bibinfo{author}{S.~K. Dwivedi}, \bibinfo{author}{C.~Arya},
\newblock \bibinfo{title}{A survey of news recommendation approaches},
\newblock in: \bibinfo{booktitle}{2016 International Conference on ICT in
  Business Industry \& Government (ICTBIG)}, \bibinfo{organization}{IEEE},
  \bibinfo{year}{2016}, pp. \bibinfo{pages}{1--6}.
%Type = Inproceedings
\bibitem[{Zheng et~al.(2018)Zheng, Zhang, Zheng, Xiang, Yuan, Xie, and
  Li}]{zheng2018drn}
\bibinfo{author}{G.~Zheng}, \bibinfo{author}{F.~Zhang},
  \bibinfo{author}{Z.~Zheng}, \bibinfo{author}{Y.~Xiang},
  \bibinfo{author}{N.~J. Yuan}, \bibinfo{author}{X.~Xie},
  \bibinfo{author}{Z.~Li},
\newblock \bibinfo{title}{Drn: A deep reinforcement learning framework for news
  recommendation},
\newblock in: \bibinfo{booktitle}{Proceedings of the 2018 world wide web
  conference}, \bibinfo{year}{2018}, pp. \bibinfo{pages}{167--176}.
%Type = Article
\bibitem[{Latifi et~al.(2021)Latifi, Mauro, and Jannach}]{LATIFI2021291}
\bibinfo{author}{S.~Latifi}, \bibinfo{author}{N.~Mauro},
  \bibinfo{author}{D.~Jannach},
\newblock \bibinfo{title}{Session-aware recommendation: A surprising quest for
  the state-of-the-art},
\newblock \bibinfo{journal}{Information Sciences} \bibinfo{volume}{573}
  (\bibinfo{year}{2021}) \bibinfo{pages}{291--315}. \URLprefix
  \url{https://www.sciencedirect.com/science/article/pii/S0020025521005089}.
  \DOIprefix\doi{https://doi.org/10.1016/j.ins.2021.05.048}.
%Type = Inproceedings
\bibitem[{Wu et~al.(2020)Wu, Qiao, Chen, Wu, Qi, Lian, Liu, Xie, Gao, Wu
  et~al.}]{wu2020mind}
\bibinfo{author}{F.~Wu}, \bibinfo{author}{Y.~Qiao}, \bibinfo{author}{J.-H.
  Chen}, \bibinfo{author}{C.~Wu}, \bibinfo{author}{T.~Qi},
  \bibinfo{author}{J.~Lian}, \bibinfo{author}{D.~Liu},
  \bibinfo{author}{X.~Xie}, \bibinfo{author}{J.~Gao}, \bibinfo{author}{W.~Wu},
  et~al.,
\newblock \bibinfo{title}{Mind: A large-scale dataset for news recommendation},
\newblock in: \bibinfo{booktitle}{Proceedings of the 58th Annual Meeting of the
  Association for Computational Linguistics}, \bibinfo{year}{2020}, pp.
  \bibinfo{pages}{3597--3606}.
%Type = Article
\bibitem[{Khan et~al.(2022)Khan, Ma, Ullah, and Polat}]{KHAN202269}
\bibinfo{author}{N.~Khan}, \bibinfo{author}{Z.~Ma}, \bibinfo{author}{A.~Ullah},
  \bibinfo{author}{K.~Polat},
\newblock \bibinfo{title}{Similarity attributed knowledge graph embedding
  enhancement for item recommendation},
\newblock \bibinfo{journal}{Information Sciences} \bibinfo{volume}{613}
  (\bibinfo{year}{2022}) \bibinfo{pages}{69--95}. \URLprefix
  \url{https://www.sciencedirect.com/science/article/pii/S0020025522010441}.
  \DOIprefix\doi{https://doi.org/10.1016/j.ins.2022.08.124}.
%Type = Article
\bibitem[{Wu et~al.(2023)Wu, Wu, Huang, and Xie}]{wu2023personalized}
\bibinfo{author}{C.~Wu}, \bibinfo{author}{F.~Wu}, \bibinfo{author}{Y.~Huang},
  \bibinfo{author}{X.~Xie},
\newblock \bibinfo{title}{Personalized news recommendation: Methods and
  challenges},
\newblock \bibinfo{journal}{ACM Transactions on Information Systems}
  \bibinfo{volume}{41} (\bibinfo{year}{2023}) \bibinfo{pages}{1--50}.
%Type = Article
\bibitem[{Pham et~al.(2023)Pham, Nguyen, Nguyen, Kozma, and Vo}]{PHAM2023105}
\bibinfo{author}{P.~Pham}, \bibinfo{author}{L.~T. Nguyen},
  \bibinfo{author}{N.~T. Nguyen}, \bibinfo{author}{R.~Kozma},
  \bibinfo{author}{B.~Vo},
\newblock \bibinfo{title}{A hierarchical fused fuzzy deep neural network with
  heterogeneous network embedding for recommendation},
\newblock \bibinfo{journal}{Information Sciences} \bibinfo{volume}{620}
  (\bibinfo{year}{2023}) \bibinfo{pages}{105--124}. \URLprefix
  \url{https://www.sciencedirect.com/science/article/pii/S0020025522013767}.
  \DOIprefix\doi{https://doi.org/10.1016/j.ins.2022.11.085}.
%Type = Inproceedings
\bibitem[{Garcin et~al.(2013)Garcin, Dimitrakakis, and
  Faltings}]{DBLP:conf/recsys/GarcinDF13}
\bibinfo{author}{F.~Garcin}, \bibinfo{author}{C.~Dimitrakakis},
  \bibinfo{author}{B.~Faltings},
\newblock \bibinfo{title}{Personalized news recommendation with context trees},
\newblock in: \bibinfo{booktitle}{RecSys}, \bibinfo{year}{2013}.
%Type = Article
\bibitem[{Zhang et~al.(2022)Zhang, Wang, Ren, Li, Deng, and
  Zhang}]{zhang2022metonr}
\bibinfo{author}{M.~Zhang}, \bibinfo{author}{G.~Wang},
  \bibinfo{author}{L.~Ren}, \bibinfo{author}{J.~Li}, \bibinfo{author}{K.~Deng},
  \bibinfo{author}{B.~Zhang},
\newblock \bibinfo{title}{Metonr: A meta explanation triplet oriented news
  recommendation model},
\newblock \bibinfo{journal}{Knowledge-Based Systems} \bibinfo{volume}{238}
  (\bibinfo{year}{2022}) \bibinfo{pages}{107922}.
%Type = Inproceedings
\bibitem[{Park et~al.(2017)Park, Lee, and Choi}]{DBLP:conf/cikm/ParkLC17}
\bibinfo{author}{K.~Park}, \bibinfo{author}{J.~Lee}, \bibinfo{author}{J.~Choi},
\newblock \bibinfo{title}{Deep neural networks for news recommendations},
\newblock in: \bibinfo{booktitle}{CIKM}, \bibinfo{year}{2017}.
%Type = Inproceedings
\bibitem[{Khattar et~al.(2018{\natexlab{a}})Khattar, Kumar, Varma, and
  Gupta}]{DBLP:conf/cikm/KhattarKV018}
\bibinfo{author}{D.~Khattar}, \bibinfo{author}{V.~Kumar},
  \bibinfo{author}{V.~Varma}, \bibinfo{author}{M.~Gupta},
\newblock \bibinfo{title}{{HRAM:} {A} hybrid recurrent attention machine for
  news recommendation},
\newblock in: \bibinfo{booktitle}{CIKM}, \bibinfo{year}{2018}{\natexlab{a}},
  pp. \bibinfo{pages}{1619--1622}.
%Type = Inproceedings
\bibitem[{Khattar et~al.(2018{\natexlab{b}})Khattar, Kumar, Varma, and
  Gupta}]{DBLP:conf/cikm/KhattarKV018a}
\bibinfo{author}{D.~Khattar}, \bibinfo{author}{V.~Kumar},
  \bibinfo{author}{V.~Varma}, \bibinfo{author}{M.~Gupta},
\newblock \bibinfo{title}{Weave{\&}rec: {A} word embedding based 3-d
  convolutional network for news recommendation},
\newblock in: \bibinfo{booktitle}{CIKM}, \bibinfo{year}{2018}{\natexlab{b}}.
%Type = Inproceedings
\bibitem[{Li et~al.(2007)Li, Li, and Tang}]{li2007flexible}
\bibinfo{author}{J.~Li}, \bibinfo{author}{J.~Li}, \bibinfo{author}{J.~Tang},
\newblock \bibinfo{title}{A flexible topic-driven framework for news
  exploration},
\newblock in: \bibinfo{booktitle}{Proceedings of KDD}, volume
  \bibinfo{volume}{2007}, \bibinfo{year}{2007}.
%Type = Inproceedings
\bibitem[{Zhang et~al.(2019)Zhang, Chen, and Ma}]{zhang2019dynamic}
\bibinfo{author}{H.~Zhang}, \bibinfo{author}{X.~Chen}, \bibinfo{author}{S.~Ma},
\newblock \bibinfo{title}{Dynamic news recommendation with hierarchical
  attention network},
\newblock in: \bibinfo{booktitle}{2019 IEEE International Conference on Data
  Mining (ICDM)}, \bibinfo{organization}{IEEE}, \bibinfo{year}{2019}, pp.
  \bibinfo{pages}{1456--1461}.
%Type = Inproceedings
\bibitem[{Vaswani et~al.(2017)Vaswani, Shazeer, Parmar, Uszkoreit, Jones,
  Gomez, Kaiser, and Polosukhin}]{vaswani2017attention}
\bibinfo{author}{A.~Vaswani}, \bibinfo{author}{N.~Shazeer},
  \bibinfo{author}{N.~Parmar}, \bibinfo{author}{J.~Uszkoreit},
  \bibinfo{author}{L.~Jones}, \bibinfo{author}{A.~N. Gomez},
  \bibinfo{author}{{\L}.~Kaiser}, \bibinfo{author}{I.~Polosukhin},
\newblock \bibinfo{title}{Attention is all you need},
\newblock in: \bibinfo{booktitle}{Advances in neural information processing
  systems}, \bibinfo{year}{2017}, pp. \bibinfo{pages}{5998--6008}.
%Type = Inproceedings
\bibitem[{Lv et~al.(2011)Lv, Moon, Kolari, Zheng, Wang, and
  Chang}]{DBLP:conf/www/LvMKZWC11}
\bibinfo{author}{Y.~Lv}, \bibinfo{author}{T.~Moon},
  \bibinfo{author}{P.~Kolari}, \bibinfo{author}{Z.~Zheng},
  \bibinfo{author}{X.~Wang}, \bibinfo{author}{Y.~Chang},
\newblock \bibinfo{title}{Learning to model relatedness for news
  recommendation},
\newblock in: \bibinfo{booktitle}{WWW}, \bibinfo{year}{2011}.
%Type = Inproceedings
\bibitem[{Hsieh et~al.(2016)Hsieh, Yang, Wei, Naaman, and
  Estrin}]{DBLP:conf/www/HsiehYWNE16}
\bibinfo{author}{C.~Hsieh}, \bibinfo{author}{L.~Yang},
  \bibinfo{author}{H.~Wei}, \bibinfo{author}{M.~Naaman},
  \bibinfo{author}{D.~Estrin},
\newblock \bibinfo{title}{Immersive recommendation: News and event
  recommendations using personal digital traces},
\newblock in: \bibinfo{booktitle}{WWW}, \bibinfo{year}{2016}.
%Type = Inproceedings
\bibitem[{Das et~al.(2007)Das, Datar, Garg, and
  Rajaram}]{DBLP:conf/www/DasDGR07}
\bibinfo{author}{A.~Das}, \bibinfo{author}{M.~Datar},
  \bibinfo{author}{A.~Garg}, \bibinfo{author}{S.~Rajaram},
\newblock \bibinfo{title}{Google news personalization: scalable online
  collaborative filtering},
\newblock in: \bibinfo{booktitle}{WWW}, \bibinfo{year}{2007}.
%Type = Inproceedings
\bibitem[{Li et~al.(2011)Li, Wang, Li, Knox, and
  Padmanabhan}]{DBLP:conf/sigir/LiWLKP11}
\bibinfo{author}{L.~Li}, \bibinfo{author}{D.~Wang}, \bibinfo{author}{T.~Li},
  \bibinfo{author}{D.~Knox}, \bibinfo{author}{B.~Padmanabhan},
\newblock \bibinfo{title}{{SCENE:} a scalable two-stage personalized news
  recommendation system},
\newblock in: \bibinfo{booktitle}{SIGIR}, \bibinfo{year}{2011}.
%Type = Article
\bibitem[{Lv et~al.(2017)Lv, Meng, and Zhang}]{DBLP:journals/ipm/LvMZ17}
\bibinfo{author}{P.~Lv}, \bibinfo{author}{X.~Meng}, \bibinfo{author}{Y.~Zhang},
\newblock \bibinfo{title}{Fere: Exploiting influence of multi-dimensional
  features resided in news domain for recommendation},
\newblock \bibinfo{journal}{Inf. Process. Manage.} \bibinfo{volume}{53}
  (\bibinfo{year}{2017}) \bibinfo{pages}{1215--1241}.
%Type = Inproceedings
\bibitem[{Wang et~al.(2018)Wang, Zhang, Xie, and Guo}]{DBLP:conf/www/WangZXG18}
\bibinfo{author}{H.~Wang}, \bibinfo{author}{F.~Zhang},
  \bibinfo{author}{X.~Xie}, \bibinfo{author}{M.~Guo},
\newblock \bibinfo{title}{{DKN:} deep knowledge-aware network for news
  recommendation},
\newblock in: \bibinfo{booktitle}{WWW}, \bibinfo{year}{2018}.
%Type = Inproceedings
\bibitem[{Zhou et~al.(2018)Zhou, Zhu, Song, Fan, Zhu, Ma, Yan, Jin, Li, and
  Gai}]{DBLP:conf/kdd/ZhouZSFZMYJLG18}
\bibinfo{author}{G.~Zhou}, \bibinfo{author}{X.~Zhu}, \bibinfo{author}{C.~Song},
  \bibinfo{author}{Y.~Fan}, \bibinfo{author}{H.~Zhu}, \bibinfo{author}{X.~Ma},
  \bibinfo{author}{Y.~Yan}, \bibinfo{author}{J.~Jin}, \bibinfo{author}{H.~Li},
  \bibinfo{author}{K.~Gai},
\newblock \bibinfo{title}{Deep interest network for click-through rate
  prediction},
\newblock in: \bibinfo{booktitle}{SIGKDD}, \bibinfo{year}{2018}, pp.
  \bibinfo{pages}{1059--1068}.
%Type = Inproceedings
\bibitem[{Lian et~al.(2018)Lian, Zhang, Xie, and Sun}]{lian2018towards}
\bibinfo{author}{J.~Lian}, \bibinfo{author}{F.~Zhang},
  \bibinfo{author}{X.~Xie}, \bibinfo{author}{G.~Sun},
\newblock \bibinfo{title}{Towards better representation learning for
  personalized news recommendation: a multi-channel deep fusion approach.},
\newblock in: \bibinfo{booktitle}{IJCAI}, \bibinfo{year}{2018}, pp.
  \bibinfo{pages}{3805--3811}.
%Type = Inproceedings
\bibitem[{Guo et~al.(2023)Guo, Yu, Shihada, and
  Zhang}]{10.1145/3543507.3583383}
\bibinfo{author}{T.~Guo}, \bibinfo{author}{L.~Yu},
  \bibinfo{author}{B.~Shihada}, \bibinfo{author}{X.~Zhang},
\newblock \bibinfo{title}{Few-shot news recommendation via cross-lingual
  transfer},
\newblock in: \bibinfo{booktitle}{Proceedings of the ACM Web Conference 2023},
  WWW '23, \bibinfo{publisher}{Association for Computing Machinery},
  \bibinfo{address}{New York, NY, USA}, \bibinfo{year}{2023}, p.
  \bibinfo{pages}{1130–1140}. \URLprefix
  \url{https://doi.org/10.1145/3543507.3583383}.
  \DOIprefix\doi{10.1145/3543507.3583383}.
%Type = Inproceedings
\bibitem[{Shi et~al.(2021)Shi, Ma, Wang, Zhang, Fang, Xu, Liu, and
  Ma}]{10.1145/3459637.3482401}
\bibinfo{author}{S.~Shi}, \bibinfo{author}{W.~Ma}, \bibinfo{author}{Z.~Wang},
  \bibinfo{author}{M.~Zhang}, \bibinfo{author}{K.~Fang},
  \bibinfo{author}{J.~Xu}, \bibinfo{author}{Y.~Liu}, \bibinfo{author}{S.~Ma},
\newblock \bibinfo{title}{Wg4rec: Modeling textual content with word graph for
  news recommendation},
\newblock in: \bibinfo{booktitle}{Proceedings of the 30th ACM International
  Conference on Information \& Knowledge Management}, CIKM '21,
  \bibinfo{publisher}{Association for Computing Machinery},
  \bibinfo{address}{New York, NY, USA}, \bibinfo{year}{2021}, p.
  \bibinfo{pages}{1651–1660}. \URLprefix
  \url{https://doi.org/10.1145/3459637.3482401}.
  \DOIprefix\doi{10.1145/3459637.3482401}.
%Type = Inproceedings
\bibitem[{Alshehri and Zhang(2022)}]{10.1145/3511808.3557335}
\bibinfo{author}{M.~A. Alshehri}, \bibinfo{author}{X.~Zhang},
\newblock \bibinfo{title}{Generative adversarial zero-shot learning for
  cold-start news recommendation},
\newblock in: \bibinfo{booktitle}{Proceedings of the 31st ACM International
  Conference on Information \& Knowledge Management}, CIKM '22,
  \bibinfo{publisher}{Association for Computing Machinery},
  \bibinfo{address}{New York, NY, USA}, \bibinfo{year}{2022}, p.
  \bibinfo{pages}{26–36}. \URLprefix
  \url{https://doi.org/10.1145/3511808.3557335}.
  \DOIprefix\doi{10.1145/3511808.3557335}.
%Type = Article
\bibitem[{Huang et~al.(2023)Huang, Luo, Liu, Zhao, and Fu}]{HUANG2023118943}
\bibinfo{author}{X.~Huang}, \bibinfo{author}{Y.~Luo}, \bibinfo{author}{L.~Liu},
  \bibinfo{author}{W.~Zhao}, \bibinfo{author}{S.~Fu},
\newblock \bibinfo{title}{Randomization is all you need: A privacy-preserving
  federated learning framework for news recommendation},
\newblock \bibinfo{journal}{Information Sciences} \bibinfo{volume}{637}
  (\bibinfo{year}{2023}) \bibinfo{pages}{118943}. \URLprefix
  \url{https://www.sciencedirect.com/science/article/pii/S0020025523005121}.
  \DOIprefix\doi{https://doi.org/10.1016/j.ins.2023.118943}.
%Type = Misc
\bibitem[{Liu et~al.(2023)Liu, Yin, Cao, Xia, Chen, and
  Zhang}]{liu2023perconet}
\bibinfo{author}{R.~Liu}, \bibinfo{author}{B.~Yin}, \bibinfo{author}{Z.~Cao},
  \bibinfo{author}{Q.~Xia}, \bibinfo{author}{Y.~Chen},
  \bibinfo{author}{D.~Zhang}, \bibinfo{title}{Perconet: News recommendation
  with explicit persona and contrastive learning}, \bibinfo{year}{2023}.
  \href{http://arxiv.org/abs/2304.07923}{{\tt arXiv:2304.07923}}.
%Type = Inproceedings
\bibitem[{Sun et~al.(2019)Sun, Liu, Wu, Pei, Lin, Ou, and
  Jiang}]{sun2019bert4rec}
\bibinfo{author}{F.~Sun}, \bibinfo{author}{J.~Liu}, \bibinfo{author}{J.~Wu},
  \bibinfo{author}{C.~Pei}, \bibinfo{author}{X.~Lin}, \bibinfo{author}{W.~Ou},
  \bibinfo{author}{P.~Jiang},
\newblock \bibinfo{title}{Bert4rec: Sequential recommendation with
  bidirectional encoder representations from transformer},
\newblock in: \bibinfo{booktitle}{Proceedings of the 28th ACM International
  Conference on Information and Knowledge Management}, \bibinfo{year}{2019},
  pp. \bibinfo{pages}{1441--1450}.
%Type = Inproceedings
\bibitem[{Wu et~al.(2019)Wu, Wu, Ge, Qi, Huang, and Xie}]{wu2019neural}
\bibinfo{author}{C.~Wu}, \bibinfo{author}{F.~Wu}, \bibinfo{author}{S.~Ge},
  \bibinfo{author}{T.~Qi}, \bibinfo{author}{Y.~Huang},
  \bibinfo{author}{X.~Xie},
\newblock \bibinfo{title}{Neural news recommendation with multi-head
  self-attention},
\newblock in: \bibinfo{booktitle}{Proceedings of the 2019 conference on
  empirical methods in natural language processing and the 9th international
  joint conference on natural language processing (EMNLP-IJCNLP)},
  \bibinfo{year}{2019}, pp. \bibinfo{pages}{6389--6394}.
%Type = Article
\bibitem[{Huang et~al.(2022)Huang, Han, Xu, and Liu}]{HUANG2022119}
\bibinfo{author}{J.~Huang}, \bibinfo{author}{Z.~Han}, \bibinfo{author}{H.~Xu},
  \bibinfo{author}{H.~Liu},
\newblock \bibinfo{title}{Adapted transformer network for news recommendation},
\newblock \bibinfo{journal}{Neurocomputing} \bibinfo{volume}{469}
  (\bibinfo{year}{2022}) \bibinfo{pages}{119--129}. \URLprefix
  \url{https://www.sciencedirect.com/science/article/pii/S0925231221015356}.
  \DOIprefix\doi{https://doi.org/10.1016/j.neucom.2021.10.049}.
%Type = Article
\bibitem[{Zhu et~al.(2022)Zhu, Cheng, Luo, Yang, Luo, Qian, and
  Zhou}]{ZHU202233}
\bibinfo{author}{P.~Zhu}, \bibinfo{author}{D.~Cheng}, \bibinfo{author}{S.~Luo},
  \bibinfo{author}{F.~Yang}, \bibinfo{author}{Y.~Luo},
  \bibinfo{author}{W.~Qian}, \bibinfo{author}{A.~Zhou},
\newblock \bibinfo{title}{Si-news: Integrating social information for news
  recommendation with attention-based graph convolutional network},
\newblock \bibinfo{journal}{Neurocomputing} \bibinfo{volume}{494}
  (\bibinfo{year}{2022}) \bibinfo{pages}{33--42}. \URLprefix
  \url{https://www.sciencedirect.com/science/article/pii/S092523122200457X}.
  \DOIprefix\doi{https://doi.org/10.1016/j.neucom.2022.04.073}.
%Type = Article
\bibitem[{Qiu et~al.(2022)Qiu, Hu, and Wu}]{10.1145/3511708}
\bibinfo{author}{Z.~Qiu}, \bibinfo{author}{Y.~Hu}, \bibinfo{author}{X.~Wu},
\newblock \bibinfo{title}{Graph neural news recommendation with user existing
  and potential interest modeling},
\newblock \bibinfo{journal}{ACM Trans. Knowl. Discov. Data}
  \bibinfo{volume}{16} (\bibinfo{year}{2022}). \URLprefix
  \url{https://doi.org/10.1145/3511708}. \DOIprefix\doi{10.1145/3511708}.
%Type = Article
\bibitem[{Huang et~al.(2023)Huang, Jin, Zhao, Liu, Lian, Tengfei, and
  Chen}]{10.1145/3555373}
\bibinfo{author}{Z.~Huang}, \bibinfo{author}{B.~Jin},
  \bibinfo{author}{H.~Zhao}, \bibinfo{author}{Q.~Liu},
  \bibinfo{author}{D.~Lian}, \bibinfo{author}{B.~Tengfei},
  \bibinfo{author}{E.~Chen},
\newblock \bibinfo{title}{Personal or general? a hybrid strategy with
  multi-factors for news recommendation},
\newblock \bibinfo{journal}{ACM Trans. Inf. Syst.} \bibinfo{volume}{41}
  (\bibinfo{year}{2023}). \URLprefix \url{https://doi.org/10.1145/3555373}.
  \DOIprefix\doi{10.1145/3555373}.
%Type = Inproceedings
\bibitem[{Yuan et~al.(2013)Yuan, Cong, Ma, Sun, and
  Magnenat-Thalmann}]{DBLP:conf/sigir/QuanYuan13}
\bibinfo{author}{Q.~Yuan}, \bibinfo{author}{G.~Cong}, \bibinfo{author}{Z.~Ma},
  \bibinfo{author}{A.~Sun}, \bibinfo{author}{N.~Magnenat-Thalmann},
\newblock \bibinfo{title}{Time-aware point-of-interest recommendation},
\newblock in: \bibinfo{booktitle}{SIGIR}, \bibinfo{year}{2013}.
%Type = Inproceedings
\bibitem[{Zhu et~al.(2017)Zhu, Li, Liao, Wang, Guan, Liu, and
  Cai}]{DBLP:conf/ijcai/ZhuLLWGLC17}
\bibinfo{author}{Y.~Zhu}, \bibinfo{author}{H.~Li}, \bibinfo{author}{Y.~Liao},
  \bibinfo{author}{B.~Wang}, \bibinfo{author}{Z.~Guan},
  \bibinfo{author}{H.~Liu}, \bibinfo{author}{D.~Cai},
\newblock \bibinfo{title}{What to do next: Modeling user behaviors by
  time-lstm},
\newblock in: \bibinfo{booktitle}{IJCAI}, \bibinfo{year}{2017}.
%Type = Inproceedings
\bibitem[{Li et~al.(2020)Li, Wang, and McAuley}]{li2020time}
\bibinfo{author}{J.~Li}, \bibinfo{author}{Y.~Wang},
  \bibinfo{author}{J.~McAuley},
\newblock \bibinfo{title}{Time interval aware self-attention for sequential
  recommendation},
\newblock in: \bibinfo{booktitle}{Proceedings of the 13th International
  Conference on Web Search and Data Mining}, \bibinfo{year}{2020}, pp.
  \bibinfo{pages}{322--330}.
%Type = Article
\bibitem[{Bamler and Mandt(2020)}]{bamler2020extreme}
\bibinfo{author}{R.~Bamler}, \bibinfo{author}{S.~Mandt},
\newblock \bibinfo{title}{Extreme classification via adversarial softmax
  approximation},
\newblock \bibinfo{journal}{arXiv preprint arXiv:2002.06298}
  (\bibinfo{year}{2020}).
%Type = Inproceedings
\bibitem[{Gulla et~al.(2017)Gulla, Zhang, Liu, {\"{O}}zg{\"{o}}bek, and
  Su}]{DBLP:conf/webi/GullaZLOS17}
\bibinfo{author}{J.~A. Gulla}, \bibinfo{author}{L.~Zhang},
  \bibinfo{author}{P.~Liu}, \bibinfo{author}{{\"{O}}.~{\"{O}}zg{\"{o}}bek},
  \bibinfo{author}{X.~Su},
\newblock \bibinfo{title}{The adressa dataset for news recommendation},
\newblock in: \bibinfo{booktitle}{Proceedings of the International Conference
  on Web Intelligence, Leipzig, Germany, August 23-26, 2017},
  \bibinfo{year}{2017}, pp. \bibinfo{pages}{1042--1048}.
%Type = Inproceedings
\bibitem[{Sheu and Li(2020)}]{10.1145/3383313.3418477}
\bibinfo{author}{H.-S. Sheu}, \bibinfo{author}{S.~Li},
\newblock \bibinfo{title}{Context-aware graph embedding for session-based news
  recommendation},
\newblock in: \bibinfo{booktitle}{Proceedings of the 14th ACM Conference on
  Recommender Systems}, RecSys '20, \bibinfo{publisher}{Association for
  Computing Machinery}, \bibinfo{address}{New York, NY, USA},
  \bibinfo{year}{2020}, p. \bibinfo{pages}{657–662}. \URLprefix
  \url{https://doi.org/10.1145/3383313.3418477}.
  \DOIprefix\doi{10.1145/3383313.3418477}.
%Type = Inproceedings
\bibitem[{Lee et~al.(2020)Lee, Oh, Seo, and Lee}]{10.1145/3340531.3411932}
\bibinfo{author}{D.~Lee}, \bibinfo{author}{B.~Oh}, \bibinfo{author}{S.~Seo},
  \bibinfo{author}{K.-H. Lee},
\newblock \bibinfo{title}{News recommendation with topic-enriched knowledge
  graphs},
\newblock in: \bibinfo{booktitle}{Proceedings of the 29th ACM International
  Conference on Information \& Knowledge Management}, CIKM '20,
  \bibinfo{publisher}{Association for Computing Machinery},
  \bibinfo{address}{New York, NY, USA}, \bibinfo{year}{2020}, p.
  \bibinfo{pages}{695–704}. \URLprefix
  \url{https://doi.org/10.1145/3340531.3411932}.
  \DOIprefix\doi{10.1145/3340531.3411932}.
%Type = Article
\bibitem[{Gao et~al.(2023)Gao, Zheng, Li, Li, Qin, Piao, Quan, Chang, Jin, He,
  and Li}]{10.1145/3568022}
\bibinfo{author}{C.~Gao}, \bibinfo{author}{Y.~Zheng}, \bibinfo{author}{N.~Li},
  \bibinfo{author}{Y.~Li}, \bibinfo{author}{Y.~Qin}, \bibinfo{author}{J.~Piao},
  \bibinfo{author}{Y.~Quan}, \bibinfo{author}{J.~Chang},
  \bibinfo{author}{D.~Jin}, \bibinfo{author}{X.~He}, \bibinfo{author}{Y.~Li},
\newblock \bibinfo{title}{A survey of graph neural networks for recommender
  systems: Challenges, methods, and directions},
\newblock \bibinfo{journal}{ACM Trans. Recomm. Syst.} \bibinfo{volume}{1}
  (\bibinfo{year}{2023}). \URLprefix \url{https://doi.org/10.1145/3568022}.
  \DOIprefix\doi{10.1145/3568022}.
%Type = Article
\bibitem[{Wu et~al.(2023)Wu, Wu, Huang, and Xie}]{10.1145/3530257}
\bibinfo{author}{C.~Wu}, \bibinfo{author}{F.~Wu}, \bibinfo{author}{Y.~Huang},
  \bibinfo{author}{X.~Xie},
\newblock \bibinfo{title}{Personalized news recommendation: Methods and
  challenges},
\newblock \bibinfo{journal}{ACM Trans. Inf. Syst.} \bibinfo{volume}{41}
  (\bibinfo{year}{2023}). \URLprefix \url{https://doi.org/10.1145/3530257}.
  \DOIprefix\doi{10.1145/3530257}.
%Type = Article
\bibitem[{Si et~al.(2023)Si, Sun, Zhang, Xu, Song, Zang, and
  Wen}]{10.1145/3582425}
\bibinfo{author}{Z.~Si}, \bibinfo{author}{Z.~Sun}, \bibinfo{author}{X.~Zhang},
  \bibinfo{author}{J.~Xu}, \bibinfo{author}{Y.~Song},
  \bibinfo{author}{X.~Zang}, \bibinfo{author}{J.-R. Wen},
\newblock \bibinfo{title}{Enhancing recommendation with search data in a causal
  learning manner},
\newblock \bibinfo{journal}{ACM Trans. Inf. Syst.} \bibinfo{volume}{41}
  (\bibinfo{year}{2023}). \URLprefix \url{https://doi.org/10.1145/3582425}.
  \DOIprefix\doi{10.1145/3582425}.
%Type = Inproceedings
\bibitem[{Yoon et~al.(2023)Yoon, Meng, Lee, and Han}]{10.1145/3543507.3583507}
\bibinfo{author}{S.~Yoon}, \bibinfo{author}{Y.~Meng}, \bibinfo{author}{D.~Lee},
  \bibinfo{author}{J.~Han},
\newblock \bibinfo{title}{Scstory: Self-supervised and continual online story
  discovery},
\newblock in: \bibinfo{booktitle}{Proceedings of the ACM Web Conference 2023},
  WWW '23, \bibinfo{publisher}{Association for Computing Machinery},
  \bibinfo{address}{New York, NY, USA}, \bibinfo{year}{2023}, p.
  \bibinfo{pages}{1853–1864}. \URLprefix
  \url{https://doi.org/10.1145/3543507.3583507}.
  \DOIprefix\doi{10.1145/3543507.3583507}.
%Type = Article
\bibitem[{Xin et~al.(2023)Xin, Yang, Wang, Ma, Ren, Luo, Shi, Chen, and
  Ren}]{10.1145/3568954}
\bibinfo{author}{X.~Xin}, \bibinfo{author}{J.~Yang}, \bibinfo{author}{H.~Wang},
  \bibinfo{author}{J.~Ma}, \bibinfo{author}{P.~Ren}, \bibinfo{author}{H.~Luo},
  \bibinfo{author}{X.~Shi}, \bibinfo{author}{Z.~Chen},
  \bibinfo{author}{Z.~Ren},
\newblock \bibinfo{title}{On the user behavior leakage from recommender system
  exposure},
\newblock \bibinfo{journal}{ACM Trans. Inf. Syst.} \bibinfo{volume}{41}
  (\bibinfo{year}{2023}). \URLprefix \url{https://doi.org/10.1145/3568954}.
  \DOIprefix\doi{10.1145/3568954}.
%Type = Inproceedings
\bibitem[{Rendle et~al.(2009)Rendle, Freudenthaler, Gantner, and
  Schmidt{-}Thieme}]{DBLP:conf/uai/RendleFGS09}
\bibinfo{author}{S.~Rendle}, \bibinfo{author}{C.~Freudenthaler},
  \bibinfo{author}{Z.~Gantner}, \bibinfo{author}{L.~Schmidt{-}Thieme},
\newblock \bibinfo{title}{{BPR:} bayesian personalized ranking from implicit
  feedback},
\newblock in: \bibinfo{booktitle}{UAI}, \bibinfo{year}{2009}.
%Type = Article
\bibitem[{Hidasi et~al.(2015)Hidasi, Karatzoglou, Baltrunas, and
  Tikk}]{DBLP:journals/corr/HidasiKBT15}
\bibinfo{author}{B.~Hidasi}, \bibinfo{author}{A.~Karatzoglou},
  \bibinfo{author}{L.~Baltrunas}, \bibinfo{author}{D.~Tikk},
\newblock \bibinfo{title}{Session-based recommendations with recurrent neural
  networks},
\newblock \bibinfo{journal}{CoRR} \bibinfo{volume}{abs/1511.06939}
  (\bibinfo{year}{2015}).
%Type = Inproceedings
\bibitem[{Tang and Wang(2018)}]{DBLP:conf/wsdm/TangW18}
\bibinfo{author}{J.~Tang}, \bibinfo{author}{K.~Wang},
\newblock \bibinfo{title}{Personalized top-n sequential recommendation via
  convolutional sequence embedding},
\newblock in: \bibinfo{booktitle}{WSDM}, \bibinfo{year}{2018}.
%Type = Inproceedings
\bibitem[{Wu et~al.(2019)Wu, Wu, An, Huang, Huang, and Xie}]{wu2019neural2}
\bibinfo{author}{C.~Wu}, \bibinfo{author}{F.~Wu}, \bibinfo{author}{M.~An},
  \bibinfo{author}{J.~Huang}, \bibinfo{author}{Y.~Huang},
  \bibinfo{author}{X.~Xie},
\newblock \bibinfo{title}{Neural news recommendation with attentive multi-view
  learning},
\newblock in: \bibinfo{booktitle}{Proceedings of the Twenty-Eighth
  International Joint Conference on Artificial Intelligence},
  \bibinfo{organization}{International Joint Conferences on Artificial
  Intelligence Organization}, \bibinfo{year}{2019}.
%Type = Inproceedings
\bibitem[{Al-Rfou et~al.(2015)Al-Rfou, Kulkarni, Perozzi, and
  Skiena}]{al2015polyglot}
\bibinfo{author}{R.~Al-Rfou}, \bibinfo{author}{V.~Kulkarni},
  \bibinfo{author}{B.~Perozzi}, \bibinfo{author}{S.~Skiena},
\newblock \bibinfo{title}{Polyglot-ner: Massive multilingual named entity
  recognition},
\newblock in: \bibinfo{booktitle}{Proceedings of the 2015 SIAM International
  Conference on Data Mining}, \bibinfo{organization}{SIAM},
  \bibinfo{year}{2015}, pp. \bibinfo{pages}{586--594}.

\end{thebibliography}

\end{document}